\documentclass[
 preprint,
 superscriptaddress,
 preprintnumbers,
 nofootinbib,
 amsmath,amssymb,
 aps, 
 prc,
 showkeys,
 floatfix,
]{revtex4-2}

\usepackage{graphicx}
\usepackage{dcolumn}
\usepackage{bm}
\usepackage{xcolor}
\usepackage{booktabs}
\usepackage[T1]{fontenc}
\usepackage{mathrsfs}
\usepackage{dsfont}
\usepackage[normalem]{ulem}
\usepackage{microtype}
\usepackage{placeins}
\usepackage[colorlinks=true,allcolors=blue]{hyperref}

\newcommand{\trento}{\texttt{T$_\mathrm{R}$ENTo} }

\begin{document}

\title{\textbf{Species-dependent viscous corrections at particlization: A novel relaxation time approximation approach}
}%

\author{I. Aguiar}
\author{T. Nunes da Silva}%
\email{Contact author: t.j.nunes@ufsc.br}
\affiliation{%
 Departamento de F\'{i}sica, Centro de Ci\^{e}ncias F\'{i}sicas e Matem\'{a}ticas, Universidade Federal de Santa Catarina, Campus Universit\'{a}rio Reitor Jo\~{a}o David Ferreira Lima, Florian\'{o}polis, Brazil, Zip Code: 88040-900
}%
\author{G. S. Denicol}
\affiliation{
 Instituto de F\'isica, Universidade Federal Fluminense,
  Av. Milton Tavares de Souza, Niter\'oi, Brazil, Zip Code: 24210-346,
}%
\author{M. Luzum}
\affiliation{
 Instituto de F\'isica, Universidade de S\~ao Paulo, R. do Mat\~ao, 1371, S\~ao Paulo, Brazil, Zip Code: 05508-090
}
\author{G. S. Rocha}
\affiliation{
 Instituto de F\'isica, Universidade Federal Fluminense,
  Av. Milton Tavares de Souza, Niter\'oi, Brazil, Zip Code: 24210-346,
}%
\affiliation{
 Department of Physics and Astronomy, Vanderbilt University, Nashville TN 37240
}%
\author{C. Shen}
\affiliation{%
Department of Physics and Astronomy, Wayne State University, Detroit, Michigan, 48201, USA
}%

\date{\today}% It is always \today, today,

\begin{abstract}
We assess the effects of a recently proposed generalized relaxation time approximation (RTA) for multi-species relativistic gases within a realistic numerical hybrid framework and study its phenomenological consequences in $p$--Pb and Pb--Pb collisions. The novel approximation introduces counter-terms to the collision kernel, allowing for momentum-dependent relaxation times $\tau_i(p)$ while preserving local energy-momentum conservation.  As a consequence, the resulting first-order viscous corrections $\delta f_i$ to the phase-space distribution functions depend explicitly on the particle species mass $m_i$.

We systematically investigate the impact of these species-dependent corrections on particle production at particlization, focusing on identified hadron yields and transverse momentum ($p_T$) spectra obtained from Cooper-Frye sampling. We find that the yields and spectra of light hadrons ($\pi, K, p$) are significantly affected, leading to modifications of relative particle yields such as the $K/\pi$ and $p/\pi$ ratios. 

We show that these effects persist, albeit with reduced magnitude, after the inclusion of the hadronic cascade stage. In contrast, the impact on inclusive charged-particle observables is strongly reduced due to compensating enhancements and suppressions among different species. This controlled deformation of identified hadron observables, which selectively modifies flavor-sensitive quantities, makes the new prescription particularly well suited for Bayesian inference, as it introduces new sensitivity directions without spoiling existing constraints.
 
Overall, our results demonstrate that species-dependent viscous corrections arising from the generalized RTA can leave significant and observable imprints on identified hadron production and relative yields, while remaining fully consistent with the successful description of bulk collective flow observables.
\end{abstract}

\maketitle

%\tableofcontents

\section{Introduction}
The study of strongly interacting matter under extreme conditions remains one of the most challenging frontiers in high-energy physics. Quantum chromodynamics (QCD), the theory governing the strong force, predicts a fundamental phase transition \cite{Gross:1973id,Politzer:1973fx} from the confined hadronic phase to a deconfined state of matter, the quark-gluon plasma (QGP) \cite{Shuryak:1978ij}, under conditions of high density and/or temperature. Despite the success of QCD in describing a wide range of phenomena, the strongly coupled nature of the theory, along with the sign problem in lattice QCD simulations \cite{deForcrand:2009zkb}, has hindered the development of a comprehensive \textit{ab initio} description of the full QCD phase diagram.

To address these questions experimentally, major efforts have been made to explore the properties of the QGP through relativistic heavy-ion collisions at facilities such as the Relativistic Heavy-Ion Collider (RHIC) and the Large Hadron Collider (LHC). These programs, which involve the collision of heavy nuclei such as Au and Pb, have provided compelling evidence that the matter produced in these collisions can be effectively described using relativistic viscous hydrodynamics. The resulting QGP behaves as a strongly-coupled fluid with a very low shear viscosity to entropy density ratio \cite{Heinz:2000bk,Arsene:2004fa,Back:2004je,Adcox:2004mh,Adams:2005dq,Muller:2012zq, ALICE:2012eyl, ATLAS:2012cix, CMS:2012qk, PHENIX:2013ktj, Heinz:2013th,Foka:2016vta, Loizides:2016tew}.

Based on these observations, the expansion of the QGP after a collision is frequently modeled using relativistic viscous hydrodynamics \cite{Israel:1976tn,Israel:1979wp}. Contemporary simulations of heavy-ion collisions adopt hydrodynamics as part of a hybrid modeling framework \cite{Petersen:2008dd,Shen:2014vra}, which integrates multiple models to account for the various phenomena that occur during a collision. This hybrid approach has proven effective in reproducing experimental data across different collision systems and center-of-mass energies.

One important step is the particlization scheme, in which the continuous degrees of freedom of the fluid at the end of hydrodynamics are translated into the discrete degrees of freedom of a hadron resonance gas, which is typically further evolved using a hadronic cascade model. It is essential to note that the particlization stage of hybrid models does not directly correspond to the physical hadronization process that occurs when the hot QGP fluid crosses the phase transition region. In hybrid models, hadronization occurs in the late stages of hydrodynamics, encoded in the equation of state used to close the set of hydrodynamic equations. Particlization takes place at the end of hydrodynamical evolution, after hadronization. There is, therefore, an implicit assumption that for some range of temperatures of the system, both descriptions, hydrodynamics and kinetic theory, are valid \cite{HotQCD:2014kol}. A particlization temperature is chosen within this range, typically from Bayesian estimation.

Particlization is typically carried out using the Cooper-Frye framework \cite{PhysRevD.10.186}, with added corrections to account for viscous non-equilibrium effects. However, the standard choice for these corrections, the Anderson-Witting relaxation time approximation (RTA), suffers from a fundamental limitation. It violates macroscopic conservation laws if the relaxation time depends on particle momentum. This forces simulations to assume constant relaxation times, ignoring the expected energy dependence of microscopic interactions.

In this work, we investigate a new prescription for particlization corrections that addresses this issue, based on the novel RTA (nRTA) approximation proposed in \cite{Rocha:2021zcw} and recently generalized to multi-species systems in \cite{Rocha:2025rkl}. This formalism introduces counter-terms that restore conservation laws even when relaxation times are momentum-dependent. We present the first full numerical implementation of this prescription in a state-of-the-art hybrid model for nuclear collision simulations. Specifically, we have studied two colliding systems, Pb-Pb and p-Pb, at different center-of-mass energies. We performed a systematic study of typical final state observables by varying the free parameter $\gamma$, which controls the power-law dependence of the relaxation time on particle energy in the new prescription for both systems. By propagating the effect through Cooper-Frye sampling, resonance content, and resonance decays, we close the connection between formal kinetic theory and experimentally relevant final-state observables. This allows, for the first time, a quantitative assessment of momentum-dependent relaxation effects on identified hadron observables within realistic heavy-ion collision simulations.

We found the relative particle yield and the identified-particles transverse momentum spectra to be especially sensitive to this parameter. This opens up the possibility of including this new prescription in future Bayesian studies, enabling a systematic assessment of species-dependent viscous effects on identified-particle yields, relative particle ratios, and baryon-to-meson enhancement patterns within a fully consistent hybrid framework.

This paper is organized as follows: in Section~\ref{sect:kinTheory}, we review the theory behind particlization models and viscous corrections; we also present how viscous corrections can be obtained using the nRTA approximation. In Section~\ref{sect:numMethods} we present the numerical setup used in our hybrid simulations of ultrarelativistic heavy-ion collisions. In Section~\ref{sect:results}, we present our results, which are summarized in Table \ref{tab:summary}, and provide a discussion of their implications. Our conclusions and future outlook are presented in Section~\ref{sect:conclusions}. Results for p-Pb collisions are presented in Appendix \ref{sect:appendix_pPb}.

\section{Particlization models and kinetic theory}
\label{sect:kinTheory}

In this section, we briefly review the relativistic Boltzmann equation and the standard relaxation time approximation to establish notation, and then summarize the generalized relaxation time approximation for multi-species systems introduced in Ref.~\cite{Rocha:2025rkl}. We emphasize the features of this formalism that are essential to the present work: the restoration of conservation laws for momentum-dependent relaxation times and the resulting species-dependent viscous corrections, which directly enter the particlization procedure discussed in Sec.~\ref{sect:numMethods}.

At the end of the hydrodynamic phase of a heavy-ion collision, hadronization takes place: the quarks and gluons within the QGP recombine into a hadron resonance gas (HRG). In the hydrodynamic regime, this process is encoded in the equation of state, which can be obtained from Lattice QCD \cite{Borsanyi:2010cj,Bazavov:2009zn}. The thermodynamic properties obtained in this formalism are compatible with the hadron resonance gas model \cite{Hagedorn:1965st,Venugopalan:1992hy,Karsch:2003vd,Huovinen:2009yb,Bazavov:2009zn,Borsanyi:2010cj,Ratti:2021ubw} at sufficiently low temperatures, and it was established that the QGP and the HRG phases are connected through a crossover \cite{Aoki:2006we}. In this way, the late stages in hybrid models for ultrarelativistic heavy-ion collisions involve the evolution of an HRG. There is an implicit assumption that, for a certain range of temperatures, both a continuum description based on hydrodynamics and a discrete description based on kinetic theory can describe the evolution of the system.  

In hybrid models, a numerical procedure known as particlization takes place within the range of temperatures where both hydrodynamics and kinetic theory are assumed to be applicable. It translates the degrees of freedom of the continuum description into those of the discrete description.
Namely, it samples discrete hadrons of a given species $i$ with phase space distribution $f_i(x,p)$ from a (typically isothermal) hypersurface $\Sigma$ constructed at the end of hydrodynamics using the Cooper-Frye formula \cite{Cooper:1974mv}:
\begin{equation}
\label{eq:cooper-frye}
    E_p \frac{dN_i}{d^3 p} = \frac{1}{\left(2\pi \hbar\right)^3} \int_\Sigma p^{\mu} d^3 \sigma_{\mu}  \, f_i(x,p), \quad i = 1, \cdots, N_{\rm spec},
\end{equation}
where $d^3 \sigma_{\mu}$ denotes the directed volume element and $N_{\rm spec}$ is the number of particle species. Throughout the present work, we will also employ the notation $f_i(x,p):= f_{{\bf p},i}$, which will also be employed for other phase-space valued functions. If the system under consideration is considered to be in local thermodynamic equilibrium, the distribution functions reduce to \cite{deGroot:80relativistic}
\begin{equation}
\label{eq:eql-f}
    f_{\text{eq},{\bf p},i} = g_i \left[ \exp\left( \frac{p^\mu_i u_{\mu}}{T}  \right) + a_i \right]^{-1},  \quad i = 1, \cdots, N_{\rm spec},
\end{equation}
where $T$ is the temperature, $u^\mu$ is the fluid velocity, $a_i = 1,$ and $-1$, respectively, for fermions and bosons, and $g_i = 2 s_{i} + 1$ is the spin degeneracy factor corresponding to spin $s_{i}$. The above distribution also considers that any chemical potential is zero. 

Since the assumption of local equilibrium cannot be upheld exactly in the context of describing the QGP evolution, the equilibrium distribution receives non-equilibrium corrections,
\begin{equation}
    f_{{\bf p},i} = f_{\text{eq},{\bf p},i} + \delta f_{{\bf p},i}.
\end{equation}
At this point, the most common strategy is to use the assumption of simultaneous validity of hydrodynamics and kinetic theory within the temperature region where particlization takes place. Then, $\delta f_{{\bf p},i}$ is related to the dissipative currents emerging from the hydrodynamic stage.

\subsection{Kinetic Theory and Hydrodynamics}

In Kinetic Theory, the dynamics of the phase space distributions for each species is provided by the relativistic Boltzmann equation, 
\begin{equation}
\label{eq:boltzmann}
\begin{aligned}
    & p^\mu_{i} \partial_\mu f_{{\bf p},i} =  C_i [f]
    \\
    &\equiv \frac{1}{2} \sum_{j,a,b=1}^{N_{\mathrm{spec}}}  \int dQ_{a} dQ'_{b} dP'_{j} \left( W_{qq' \to pp'}^{ab \to ij} 
f_{{\bf q}, a} f_{{\bf q}', b} \tilde{f}_{{\bf p}, i}
\tilde{f}_{{\bf p}', j}
-
W_{pp' \to qq'}^{ij \to ab}
f_{{\bf p}, i} f_{{\bf p}', j} \tilde{f}_{{\bf q}, a}
\tilde{f}_{{\bf q}', b}
\right),
\end{aligned}
\end{equation}
where $C_i[f]$ denotes the collision term, which encodes the various possible interactions among different particle species. Above, we display its form for two-to-two processes. Besides, we define $\tilde{f}_{{\bf p},i}  = 1 - (a_i/g_i) f_{{\bf p},i}$ and the integral measure for on-shell particles, $dP_{j} = d^{3}p_{j}/[(2 \pi)^{3} p^{0}_{j}]$, and $W_{qq' \to pp'}^{ab \to ij} \propto \delta^{(4)}(q_{i} + q_{j}' - p_{a} - p_{b}')$ is the transition rate, which is also proportional to the cross section for the corresponding processes \cite{deGroot:80relativistic} and the delta function enforces that the processes conserve the particles' four-momenta. Close to the local equilibrium configuration \eqref{eq:eql-f}, it is justifiable to linearize the collision term around local equilibrium. Then,
\begin{equation}
    \begin{split}
    \label{eq:lin-coll-t}
        C_i[f] &\simeq \frac{1}{2} \sum_{j, a, b} \int dQ_a dQ_b' dP_i' W_{qq' \to pp'}^{ab \to ij} f_{\text{eq}, {\bf p}, j} f_{\text{eq}, {\bf p}', i} \tilde{f}_{\text{eq}, {\bf q}, a} \tilde{f}_{\text{eq}, {\bf q'}, b}  \big( \phi_{{\bf q}, a} + \phi_{{\bf q}', b} - \phi_{{\bf p}, j} - \phi_{{\bf p}', i} \big) \\
        & \equiv f_{\text{eq}, {\bf p}, i} \hat{L} \phi_{{\bf p}, i},
    \end{split}
\end{equation}
which defines the linearized collision operator $\hat{L} \phi_{{\bf p},i}$, which depends on all of the individual deviation functions, $\phi_{{\bf p},i}= \delta f_ {{\bf p},i} /(f_{\text{eq}, {\bf p},i} \tilde{f}_{\text{eq}, {\bf p},i})$, $i = 1, \cdots, N_{\mathrm{spec}}$. We also employ the following shorthand notation for the summation over particle species $\sum_{i} := \sum_{i=1}^{N_{\rm spec}}$, and we remark that all summations over species will be explicitly written.

In the hydrodynamic regime, local conservation laws play a prominent role. In particular, for systems where net charge effects can be neglected, the local conservation laws for energy and momentum are the main dynamical equations,
\begin{equation}
\label{eq:consv-emt}
\begin{aligned}
    \partial_{\mu}T^{\mu \nu} &= 0,    
\end{aligned}
\end{equation}
where $T^{\mu \nu}$ is the energy-momentum tensor. In terms of the phase-space distribution, we can express it as, \cite{deGroot:80relativistic} 
\begin{equation}
\label{eq:Tmunu-KT}
\begin{aligned}
   T^{\mu \nu} & \equiv \sum_{i} \int dP_{i} \ p^{\mu}_{i} p^{\nu}_{i} f_{{\bf p}, i},    
\end{aligned}    
\end{equation}
and indeed Eq.~\eqref{eq:consv-emt} can be derived from the Boltzmann equation \eqref{eq:boltzmann}. 

In non-equilibrium configurations, Landau matching conditions provide meaning for the thermodynamic fields $T, u^{\mu}$. Then, $T$ is defined so that the total energy density follows the equilibrium equation of state, and $u^\mu$ is defined as the time-like normalized eigenvector of the energy-momentum tensor \cite{LandauLifshitzFluids}. In this case, the energy-momentum tensor can be decomposed as
\begin{equation}
\label{eq:tmunu-comps}
\begin{aligned}
T^{\mu \nu}  &= \varepsilon u^{\mu} u^{\nu} - (P+\Pi) \Delta^{\mu \nu} + \pi^{\mu \nu},    
\\
\varepsilon &= \sum_{i}\int dP_{i} E_{{\bf p},i}^{2} f_{\text{eq}, {\bf p},i}, 
\quad 
P = - \frac{1}{3} \sum_{i} \int dP_{i} \left( \Delta_{\mu \nu} p_{i}^{\mu} p_{i}^{\nu} \right) f_{\text{eq}, {\bf p},i},  \\
\Pi &= - \frac{1}{3} \sum_{i} \int dP_{i} \ \left( \Delta_{\mu \nu} p_{i}^{\mu} p_{i}^{\nu} \right) \delta f_{{\bf p},i}, 
\quad 
\pi^{\mu \nu} = \sum_{i} \int dP_{i} p^{\langle \mu}_{i} p^{\nu \rangle}_{i} \delta f_{{\bf p},i},
\end{aligned}    
\end{equation}
where we introduce the energy density $\varepsilon$, the equilibrium pressure, which follows the equation of state $P = P(\varepsilon)$; the bulk viscous pressure, $\Pi$; and the shear-stress tensor, $\pi^{\mu \nu}$. In the phase space integrals, $p^{\langle \mu}_{i} p^{\nu \rangle}_{i} = \Delta^{\mu \nu}_{\ \  \alpha \beta} p^{\alpha}_{i} p^{\beta}_{i}$, expressed in terms of the double-symmetric and traceless tensor $\Delta^{\mu \nu \alpha \beta} = (1/2) \left( \Delta^{\mu \alpha} \Delta^{\nu \beta} + \Delta^{\mu \beta} \Delta^{\nu \alpha} \right) - (1/3) \Delta^{\mu \nu} \Delta^{\alpha \beta}$. 

From Eq.~\eqref{eq:Tmunu-KT}, we readily note that many different functional forms for $f_{{\bf p}, i}$ can lead to the same energy-momentum tensor. Thus, in general non-equilibrium configurations, the problem of obtaining the form of the $f_{{\bf p},i}$ from a given $T^{\mu \nu}$ is ill-defined. Near the local equilibrium state, however, it is possible to devise prescriptions to express the distribution functions $f_{{\bf p}, i}$ in terms of the dissipative currents. Indeed, sufficiently close to local equilibrium, it is reasonable to expect that the corrections $\delta f_{{\bf p},i}$ are linear in the dissipative currents. One way to obtain the form of the corrections $\delta f_{{\bf p},i}$ in terms of hydrodynamic variables is through the Chapman-Enskog expansion \cite{chapman1916vi,enskog1917kinetische}. Summarily, it implements in Kinetic Theory the wide separation between microscopic and macroscopic degrees of freedom expected in the hydrodynamic regime through a gradient expansion. The zeroth-order result is the local equilibrium distribution \eqref{eq:eql-f} and the first-order solution (at vanishing chemical potential) reads \cite{deGroot:80relativistic,cercignani:02relativistic,Denicol:2021,Rocha:2022ind,Rocha:2023ilf}
\begin{equation}
\label{eq:phi-ST}
\begin{aligned}
&
\phi_{{\bf p}, i} = \frac{\delta f_{{\bf p}, i}}{f_{\text{eq}, {\bf p},i} \tilde{f}_{\text{eq}, {\bf p},i}} \simeq 
S_{{\bf p},i} 
\theta 
+
T_{{\bf p},i} p^{\langle \mu}_{i}p^{\nu \rangle}_{i} \sigma_{\mu \nu}, 
\quad
i = 1, \cdots, N_{\rm spec},
\end{aligned}    
\end{equation}
where $\theta = \partial_{\mu}u^{\mu}$, $\sigma_{\mu \nu} = \Delta^{\mu \nu \alpha \beta}\partial_{\alpha}u_{\beta}$. In general, the momentum space functions $S_{{\bf p},i}$ and $T_{{\bf p},i}$ can be expressed in terms of the linearized collision term \eqref{eq:lin-coll-t}. From the above expression, we see from the definitions \eqref{eq:tmunu-comps} that the dissipative currents obey constitutive relations characteristic of Navier-Stokes theory, 
\begin{subequations}
\label{eq:def-coeffs}
\begin{align}
\Pi \simeq - \zeta \theta,  \quad
\pi^{\mu \nu} \simeq  2 \eta \sigma^{\mu \nu},
\end{align}
\end{subequations}
where $\zeta$ is the bulk viscosity and $\eta$ is the shear viscosity, which are non-trivial functions of the temperature. In this regime, we can employ the constitutive relations \eqref{eq:def-coeffs} in Eq.~\eqref{eq:phi-ST} to derive
\begin{subequations}
\label{eq:CF1}
\begin{align}
\label{eq:CF1-p}
\phi_{{\bf p},i}
& =
 \phi_{{\bf p},i} \big\vert_{\mathrm{bulk}}
+
 \phi_{{\bf p},i} \big\vert_{\mathrm{shear}}  ,
\\
\label{eq:CF1-bulk}
\phi_{{\bf p},i} \big\vert_{\mathrm{bulk}}
&=
- \frac{S_{{\bf p},i}}{\zeta} \Pi,
\\
\label{eq:CF1-shear}
\phi_{{\bf p},i} \big\vert_{\mathrm{shear}}  
&=
\frac{T_{{\bf p},i}}{2 \eta} p^{\langle \mu}_{i}  p^{\nu \rangle}_{i} \pi_{\mu \nu},
\end{align}    
\end{subequations}
where $\phi_{{\bf p},i} \big\vert_{\mathrm{bulk}}$ and $\phi_{{\bf p},i} \big\vert_{\mathrm{shear}}$ are referred to as the bulk and shear sectors of the particle distribution function. The above expression is widely employed in particlization models \cite{McNelis:2019auj,Shen:2014vra}. With rare exceptions \cite{Denicol:2022bsq}, it is in general not possible to obtain analytical expressions for transport coefficients or the functions $S_{{\bf p},i}$ and $T_{{\bf p},i}$, when employing the collision term or its linearized counterpart. This motivates the use of phenomenological approximations, such as the relaxation time approximation, which we discuss next.

\subsection{RTA non-equilibrium corrections}

One of the most widely used phenomenological approximations to the collision term is the relaxation time approximation (RTA). The traditional RTA for multiple particle species is based on the Anderson-Witting \cite{ANDERSON1974489} ansatz,
\begin{equation}
\label{eq:AW-RTA}
    f_{\text{eq}, {\bf p}, i} \hat{L} \phi_{{\bf p},i} \simeq-\frac{E_{{\bf p},i}}{\tau_{R}} \delta f_{{\bf p},i}=-\frac{E_{{\bf p}, i}}{\tau_{R}} f_{\text{eq}, {\bf p},i} \tilde{f}_{\text{eq}, {\bf p}, i} \phi_{{\bf p},i}.
\end{equation}
which implements the idea that deviations from local equilibrium decay exponentially within a characteristic timescale. The above expression is nevertheless restrictive, since it is compatible with the macroscopic local conservation laws of energy and momentum \eqref{eq:consv-emt} if $\tau_{R}$ does not depend on the particle species (no $i$ dependence) and if $\tau_{R}$ does not depend on the particle four-momenta (no $p^{\mu}_{i}$ dependence $\forall i = 1, \cdots N_{\rm spec}$). These dependencies are nevertheless reasonable because $\tau_{R}$ summarizes the microscopic interactions, and in general, these interactions can lead to a system where the equilibration timescale can vary from species to species. Moreover, within a single particle species, particles with different momenta can also have different equilibration timescales (see Refs.~\cite{Rocha:2021zcw,Rocha:2025rkl,Rocha:2022crt,Dash:2021ibx,Denicol:2022bsq} for related discussions). Moreover, compatibility with the local conservation laws is only possible if Landau matching conditions are employed \cite{Rocha:2025rkl}.

To circumvent the inconsistency with the local conservation laws, the following ansatz was put forward for multi-species systems in Ref.~\cite{Rocha:2025rkl}
\begin{subequations}
    \begin{align}
    \label{eq:nRTA}
        f_{\text{eq}, {\bf p},i} \hat{L}\phi_{{\bf p},i} \ \simeq &- \frac{E_{{\bf p},i}}{\tau_{R{\bf p},i}} f_{\text{eq}, {\bf p},i} \tilde{f}_{\text{eq}, {\bf p},i} \Bigg [ \phi_{{\bf p},i} - \frac{\sum _j \left \langle \phi_{{\bf p},j}, E_{{\bf p},j} \right \rangle}{\sum _j \left \langle E_{{\bf p},j}, E_{{\bf p},j} \right \rangle} E_{{\bf p},i}
        -  \frac{\sum _j \left \langle \phi_{{\bf p},j}, p_j ^{\langle \mu \rangle} \right \rangle}{\frac{1}{3}\sum _j \left \langle p_j ^{\langle \nu \rangle}, p_{j\langle \nu \rangle} \right \rangle} p_{i\langle \mu \rangle} \Bigg ],\\
    &   
    \label{eq:nRTA-inner-prod}
        \left\langle G_{{\bf p}, j}, H_{{\bf p}, j}\right\rangle :=
        \int d P_j \frac{E_{{\bf p}, j}}{\tau_{R{\bf p}, j}} G_{{\bf p}, j} H_{{\bf p}, j} f_{\text{eq}, {\bf p}, j} \tilde{f}_{\text{eq}, {\bf p}, j} ,   
    \end{align}
\end{subequations}
where the RTA timescale $\tau_{R{\bf p},i}$ can now depend arbitrarily on the particle species and its momentum without violating the local conservation laws. Moreover, the ansatz \eqref{eq:nRTA} reduces to the Anderson-Witting ansatz \eqref{eq:AW-RTA} when $\tau_{R,i}$ is constant with respect to the particle species and with respect to the momentum of each particle. The remaining counter-terms are constructed from an orthogonal basis constructed from the components of the four-momentum, which are the microscopically conserved quantities, see Ref.~\cite{Rocha:2025rkl} for a more complete discussion. 

Also in Ref.~\cite{Rocha:2025rkl}, analytical formulas for the first-order Chapman-Enskog expansion for the phase space distribution transport coefficients for hadron-resonance-gas have been derived considering the following parametrization of the relaxation times in terms of the energy
\begin{equation}
\label{eq:polyn-dep-tR}
\tau_{R{\bf p},i} = t_{R} \left( \frac{E_{{\bf p},i}}{T} \right)^{\gamma}, 
\end{equation} 
where $t_{R}$ denotes a timescale that does not depend on particle species nor on their momentum. Moreover, the phenomenological parameter $\gamma$, which shall be considered the same for all particle species, summarizes the effect of the microscopic interactions. For $\gamma = 0$, we recover the Anderson-Witting ansatz, which can only be employed consistently with local conservation laws in the Landau frame. The case $\gamma = 1$ models well the near-equilibrium behavior emerging from scalar particles with a quartic interaction \cite{Calzetta:1986cq,Denicol:2022bsq}. The counter-terms in \eqref{eq:nRTA} ensure consistency with local conservation laws, independently of the hydrodynamic frame \cite{Rocha:2021zcw}, and, as we shall see, they give rise to new species-dependent structures in the Cooper-Frye formalism. 

The final result \cite{Rocha:2025rkl} for the first-order Chapman-Enskog expansion employing the ansatz \eqref{eq:nRTA} and the relaxation time parametrization \eqref{eq:polyn-dep-tR} can be cast in a form similar to that of Eq.~\eqref{eq:CF1}  
\begin{equation}
\begin{aligned}
&
\phi_{{\bf p},i}
& =
\phi_{{\bf p},i} \big\vert_{\mathrm{bulk}}
+
\phi_{{\bf p},i} \big\vert_{\mathrm{shear}}  ,
\end{aligned}    
\end{equation}
where the bulk sector contribution reads\footnote{Expressions for generic parametrizations of $\tau_{R {\bf p}, i}$ as a function of momentum (other than \eqref{eq:polyn-dep-tR}) can be found in Ref.~\cite{Rocha:2025rkl}}
\begin{subequations}
\label{eq:phi_bulk}
\begin{align}
\phi_{{\bf p},i} \big\vert_{\mathrm{bulk}}  
&=
\left[
\varphi(T) \frac{E_{{\bf p},i}}{T}
- 
\frac{m_{i}^{2}}{3 T^{2}} \left( \frac{E_{{\bf p},i}}{T} \right)^{\gamma - 1}  
+ 
\left( \frac{1}{3} - c_{s}^{2} \right) \left( \frac{E_{{\bf p},i}}{T} \right)^{\gamma + 1} 
 \right] \frac{\Pi}{B_{\Pi}},
\\ 
\label{eq:varphi}
\varphi(T)
&:=
\frac{1}{T^{\gamma}} \frac{1}{\mathcal{J}_{3,0}} \left(c_{s}^{2} \mathcal{J}_{\gamma+3,0} 
-  
\mathcal{J}_{\gamma+3,1} \right),
\\
\label{eq:B_PI}
B_{\Pi} & := \frac{\zeta}{t_{R}}
=
\frac{1}{T^{\gamma+1}}
\left(- c_{s}^{2} \mathcal{J}_{\gamma+3,1} + \frac{5}{3} \mathcal{J}_{\gamma+3,2} \right) 
+
\frac{1}{T^{\gamma+1}}\frac{\mathcal{J}_{3,1}}{\mathcal{J}_{3,0}}
\left(c_{s}^{2} \mathcal{J}_{\gamma+3,0} 
-  
\mathcal{J}_{\gamma+3,1} \right),
\\
\label{eq:cs2}
c_{s}^{2} & := \frac{\partial P}{\partial \varepsilon} = \frac{\mathcal{J}_{3,1}}{\mathcal{J}_{3,0}}.
\end{align}
\end{subequations}
On the other hand, the shear sector expression reads
\begin{subequations}
\label{eq:phi_shear}
\begin{align}
\phi_{{\bf p},i} \big\vert_{\mathrm{shear}}  
& =
\frac{1}{T^{2}} \left( \frac{E_{{\bf p},i}}{T} \right)^{\gamma-1}   
 p^{\langle \mu}_{i}  p^{\nu \rangle}_{i} \frac{\pi_{\mu \nu}}{2 B_{\pi}},
 \\\
\label{eq:B_shear} 
B_{\pi} & = \frac{\eta}{t_{R}}
=
\frac{1}{T^{\gamma+1}}\mathcal{J}_{\gamma+3,2}.
\end{align}        
\end{subequations}
In equations \eqref{eq:varphi}, \eqref{eq:B_PI}, \eqref{eq:cs2} and \eqref{eq:B_shear}, we employed the following notation for thermodynamic integrals
\begin{subequations}
\label{eq:thermodynamic_integrals}
\begin{align}
\label{eq:thermodynamic_integrals-J}
\mathcal{J}_{n,q} &= \sum_{i}  J_{n,q}^{(i)},
 \\
 \label{eq:thermodynamic_integrals-Ji}
 J_{n,q}^{(i)} &= \frac{1}{(2q+1)!!} \int dP_{i} \left( -\Delta_{\lambda \sigma}p_i^{\lambda} p_i^{\sigma} \right)^{q} E_{\mathbf{p},i}^{n-2q} f_{\text{eq}, {\bf p},i}\tilde{f}_{\text{eq}, {\bf p},i}. 
\end{align}
\end{subequations}
We note that, in the above expressions, the relative deviation from equilibrium for a given particle species, $\phi_{{\bf p},i}$, is expressed in terms of coefficients that reflect properties of the system as a whole. This dependence on global properties is encoded in the coefficients $\varphi(T)$, $c_{s}^{2}$, $B_{\Pi}$, and $B_{\pi}$. Plots for these quantities for the hadron-resonance gas can be seen in Fig.~1 of Ref.~\cite{Rocha:2025rkl} and are omitted here for the sake of brevity. The $\varphi(T)$ term is a novel contribution, which does not have a counterpart in the traditional RTA formulation, that arises from the momentum conservation counter-terms. It is a non-monotonic function of the temperature, whose range of values increases with the nRTA parameter $\gamma$ in Eq.~\eqref{eq:polyn-dep-tR}. We also note that $\gamma$ also influences the coefficients $B_{\Pi}$, and $B_{\pi}$, but not $c_{s}^{2}$, that is a purely thermodynamic property. Additionally, we remark that only in the case $\gamma = 0$, the timescale $t_{R}$ coincides with both bulk and shear relaxation times. Also, in this limit, $\varphi(T) = 0$ and the traditional form of the RTA result for Cooper-Frye is recovered.

\begin{figure}[t!]
    \centering
    % left, bottom, right, and top
    \includegraphics[trim={0.38cm 0.4cm 0.35cm 0.4cm}, clip, width=0.49\linewidth]{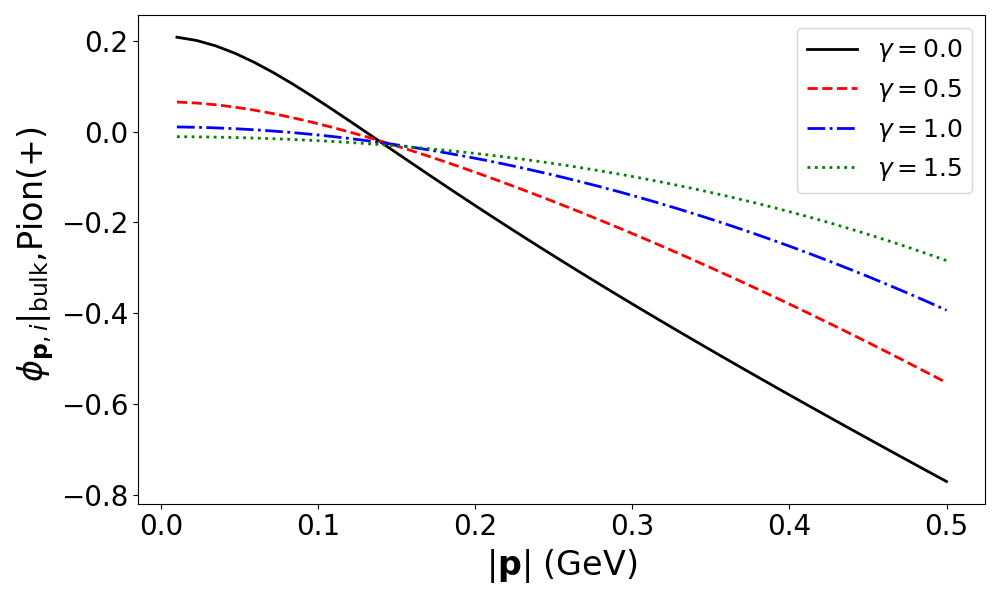} 
    \includegraphics[trim={0.4cm 0.4cm 0.35cm 0.4cm}, clip, width=0.49\linewidth]{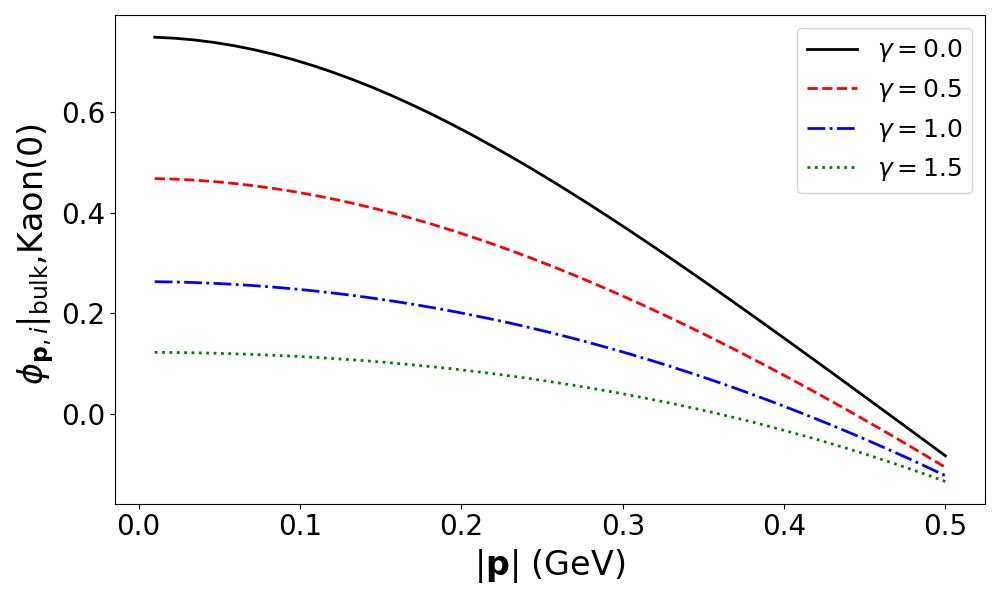}
    \includegraphics[trim={0.4cm 0.4cm 0.35cm 0.38cm}, clip, width=0.49\linewidth]{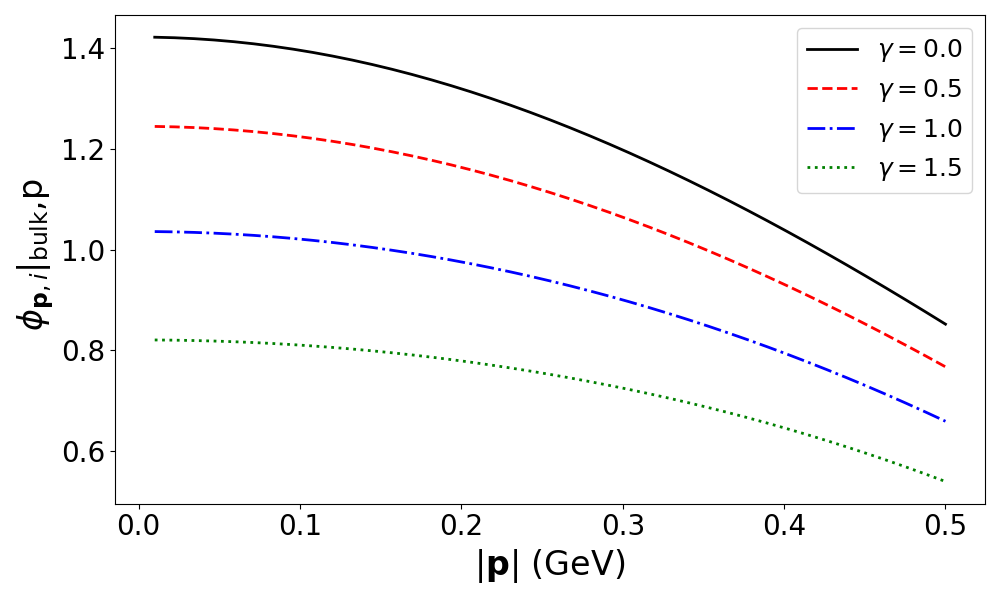}
    \caption{The bulk viscous correction $\phi_{\mathbf{p},i} |_{\mathrm{bulk}}$ (considering the typical value of  $\Pi/(\varepsilon + P) = - 0.08$) as a function of the magnitude of momentum for pions, kaons, and protons at a freeze-out temperature of $T = 0.15$ GeV. The calculation assumes $\gamma =$ 0, 0.5, 1.0, and 1.5. Note the significant splitting between particle species at low momentum, driven by the explicit mass dependence $m_i$ in the correction terms (see Eq. \eqref{eq:phi_bulk}).}
    \label{fig:nch-multiplicity}
\end{figure}

From expressions \eqref{eq:phi_bulk} and \eqref{eq:phi_shear}, we readily see that the effects on shear sector deviation function, $\phi_{{\bf p},i} \big\vert_{\mathrm{shear}}$, stem from  $B_{\pi}$ and $(E_{{\bf p},i}/T)^{\gamma - 1}$. The most non-trivial effects of the new ansatz come from the bulk sector deviation function, $\phi_{{\bf p},i} \big\vert_{\mathrm{bulk}}$. In order to assess them, we consider a hadron-resonance with the UrQMD particle listing \cite{Bass:1998ca,Bleicher:1999xi}. Then we plot $\phi_{{\bf p},i} \big\vert_{\mathrm{bulk}}$ as a function of the magnitude of momentum for pions, kaons, and protons at $T = 0.15$ GeV and different values of the RTA parameter $\gamma$ in Fig.~\ref{fig:nch-multiplicity}. We emphasize that the continuous black curves for $\gamma = 0$ amount to the traditional Anderson-Witting RTA results. We also remark that different particle species with the same mass share the same results for $\phi_{{\bf p},i} \big\vert_{\mathrm{bulk}}$. Thus, for instance, predictions for all pion species, $\pi^{0, \pm}$, are the same.
In the panels of Fig.~\ref{fig:nch-multiplicity} below, we see that in general $\phi_{{\bf p},i} \big\vert_{\mathrm{bulk}}$ decreases monotonously in the range of momenta analyzed, but the range of values of the function vary significantly with the particle species and with the RTA parameter $\gamma$. For instance, the behavior at low momentum for example is mainly driven by the different species masses in the correction terms, thus generating a splitting between species in this regime. We also notice that, for pions the curves bunch and cross each other near $ \vert {\bf p} \vert \sim 0.15$ GeV. Similar structures emerge for the other particle species at other values of $\vert {\bf p} \vert$. 

\section{Numerical methods}
\label{sect:numMethods}

In the upcoming sections, we shall assess the effect of employing Eqs.~\eqref{eq:phi_bulk} and \eqref{eq:phi_shear} as viscous corrections in the Cooper-Frye formula Eq.~\eqref{eq:cooper-frye}. The numerical analysis involved simulations of collisions for a large and a small system, namely Pb-Pb at center-of-mass energy $\sqrt{s_\text{NN}} = 2.76$ TeV and p-Pb at $\sqrt{s_\text{NN}} = 5.02$ TeV. For both systems, we used a hybrid chain which comprised:
\begin{itemize}
    \item \trento as an initial profile generator \cite{Moreland:2014oya};
    \item free-streaming pre-equilibrium dynamics \cite{Broniowski:2008qk,Liu:2015nwa};
    \item MUSIC as the hydrodynamics solver \cite{Schenke:2010nt, Schenke:2011bn,Ryu:2015vwa,Paquet:2015lta};
    \item iSS for the particlization stage \cite{Shen:2014vra};
    \item UrQMD as for the hadronic cascade \cite{Bass:1998ca, Bleicher:1999xi}.
\end{itemize}
The parameters for \textsc{TRENTo} and the hydrodynamic transport coefficients were taken from existing Bayesian analysis results. For Pb--Pb simulations, we used the maximum a posteriori (MAP) values obtained by the JETSCAPE Collaboration \cite{JETSCAPE:2020mzn}, which were calibrated using the traditional RTA $\delta f$ correction. For p--Pb simulations, we employed the MAP values obtained by the Duke group using the PTB $\delta f$ correction \cite{Moreland:2018gsh}. We emphasize that no parameters are refitted in this work. Instead, our goal is to isolate the effect of the particlization corrections by comparing the new prescription to previously validated parameter sets. A full Bayesian refit incorporating the generalized RTA will be left for future work.

For Pb-Pb simulations, the free-streaming stage uses a fixed free-streaming velocity $v_\text{fs} = c$ and a fluctuating free-streaming time, following the approach used in the aforementioned JETSCAPE work. The p-Pb simulations were performed with a fixed free-streaming velocity $v_\text{fs} = c$ and a fixed free-streaming time, following the setup by the Duke group. 

In our numerical implementation, tables for the nRTA viscous correction coefficients $B_{\pi}$, $B_{\Pi}$, $c_s ^2$, $\varphi(T)$ are calculated for different values of the phenomenological parameter $\gamma$, namely $\gamma = 0.0, \ 0.5, \ 1.0  \text{ and } 1.5$. These are then used to sample the different particle species and their momentum from each cell of the particlization hypersurface. 

The sampling routine based on the Cooper-Frye formula estimates the yield of each particle species in a hypersurface cell, 
\begin{equation}
    N_{\text{total}, i} = n_{\text{eq},i} + \delta N_i \ ,
\end{equation}
where $n_{\text{eq},i}$ is the $i$-th species equilibrium particle density, which can be expressed in terms of a thermodynamic integral as $ n_{\text{eq},i} = J_{1,0}^{(i)}$ and $\delta N_i$ is the corresponding nRTA viscous contribution given by % multiplied by the cell volume, where}
\begin{equation}
   \begin{split}
       \delta N_i := \int dP_{i} E_{{\bf p},i} \delta f_{{\bf p},i}  = \Bigg [ \frac{\varphi(T)}{T} J^{(i)}_{2,0}
- 
\frac{m_{i}^{2}}{3 T^{\gamma + 1}} J^{(i)}_{\gamma,0}  
+ 
\left( \frac{1}{3} - c_{s}^{2} \right) 
\frac{J^{(i)}_{\gamma+2,0}}{T^{\gamma+1}}
 \Bigg ] \frac{\Pi}{B_{\Pi}} .
   \end{split}
\end{equation}
The total yield estimator $N_{\text{total}, i}$ is regulated so that $N_{\text{yield},i} = \mathrm{max} (0, \ N_{\text{total}, i})$, where $N_{\text{yield},i}$ is the total non-negative yield for the $i$-th particle species. Negative contributions to the total yield are localized to small regions of phase space where viscous corrections become large; the above prescription is therefore applied only as a technical regulator and is consistent with standard implementations of particlization in existing hybrid frameworks. 

For each species in a given hypersurface cell, particles with a given momentum are then sampled from the corrected distribution function, incorporating the novel viscous correction effects also in the momentum distribution of sampled particles.

\begin{figure}[t!]
    \centering
    \includegraphics[width=0.49\linewidth]{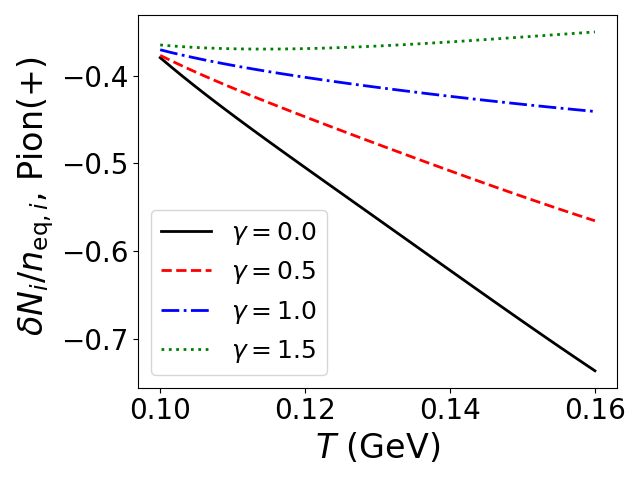} 
    \includegraphics[width=0.49\linewidth]{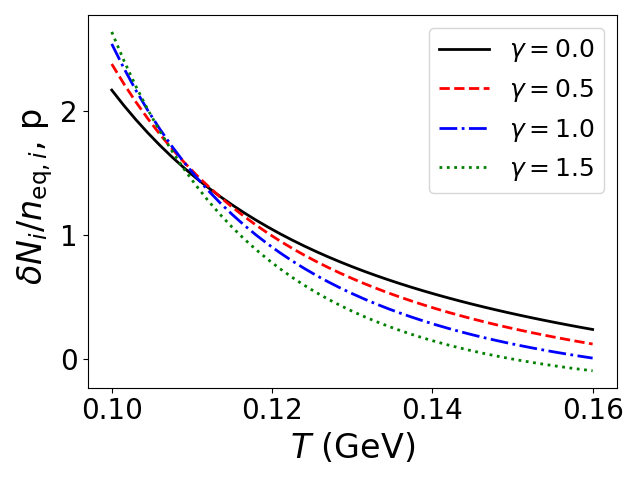}
    \includegraphics[width=0.49\linewidth]{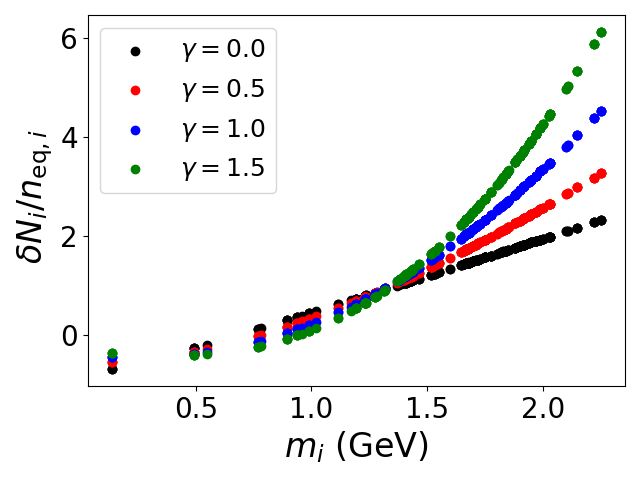}
    \includegraphics[width=0.49\linewidth]{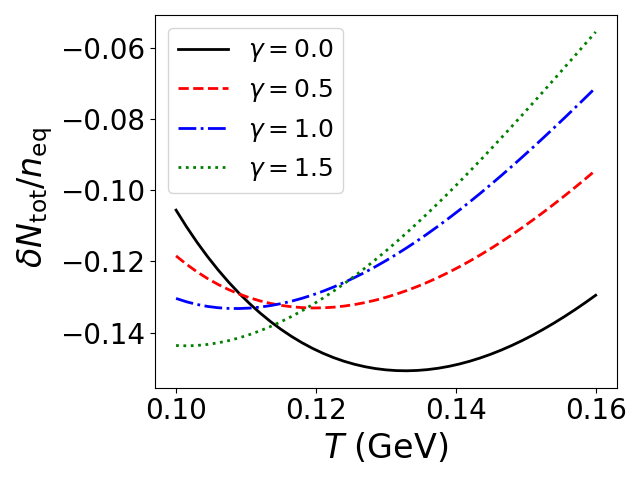}
    \caption{Upper panels -- Relative viscous correction to particle yield, $\delta N_i/n_{{\rm eq,} i}$ ($i =$ pion, proton, kaon) as a function of temperature. Lower left panel -- $\delta N_i/n_{{\rm eq,} i}$ as a function of the species mass for all species considered at $T = 0
    .15$ GeV. Lower right panel -- $\delta N_{tot}/ n_{{\rm eq}} $ as a function of temperature.  For all panels, the typical value of  $\Pi/(\varepsilon + P) = - 0.08$ is considered. }
    \label{fig:theory-deltaNs}
\end{figure}

In Fig.~\ref{fig:theory-deltaNs}, we assess the effects of the new RTA ansatz in the viscous correction to the particle densities. Here, we also consider the UrQMD particle listings. In the upper panels of Fig.~\ref{fig:theory-deltaNs}, we plot the relative viscous correction $\delta N_i/n_{{\rm eq,} i}$ as a function of temperature for pions and protons. We see that for pions, the viscous corrections tend to deplete this species in the range of temperature analyzed. This depletion is attenuated in intensity as the RTA parameter $\gamma$ increases, but increases in intensity as the temperature rises. For protons, on the other hand, the viscous correction enhances the density. Its intensity, however, decreases as a function of temperature. The enhancement behavior is non-monotonic with respect to $\gamma$ across different temperatures. 

In order to assess the species behavior in more detail, we plot $\delta N_i/n_{{\rm eq,} i}$ as a function of the species mass at $T=0.15$ GeV for different values of $\gamma$ in the lower left panel of Fig.~\ref{fig:theory-deltaNs}. We see a non-monotonic behavior with the RTA parameter $\gamma$ for fixed species. For instance, for pions ($m_{i} = 0.138$ GeV) $\delta N_i/n_{{\rm eq,} i}$ increases with $\gamma$, for kaons ($m_{i} = 0.494$ GeV) and protons ($m_{i} = 0.938$ GeV) $\delta N_i/n_{{\rm eq,} i}$ decreases with $\gamma$ and for $\Omega$ baryons ($m_{i} = 1.672$ GeV) $\delta N_i/n_{{\rm eq,} i}$ increases with $\gamma$. In the lower right panel, Fig.~\ref{fig:theory-deltaNs}, we assess predictions for global effects by plotting the relative total viscous correction to particle densities $\delta N_{tot}/ n_{{\rm eq}}$, where 
\begin{equation}
\begin{aligned}
& \delta N_{\rm tot} = \sum_{i}\delta N_i, \quad  n_{{\rm eq}} = \sum_{i} n_{{\rm eq,} i}.  
\end{aligned}
\end{equation}
Then, we see that $\delta N_{tot}/ n_{{\rm eq}}$ is typically negative within the temperatures analyzed, the total particle yield should be depleted. The intensity of this depletion is typically smaller than the typical values of the individual relative yields $\delta N_i/n_{{\rm eq,} i}$. This intensity is also non-monotonic with $\gamma$ at constant temperature: it decreases with $\gamma$ around $T = 0.10$ GeV, but it increases with $\gamma$ at $T = 0.15$ GeV.

\section{Results} 
\label{sect:results}

In this section, we first isolate the effects of the generalized relaxation time approximation at the particlization stage, and then assess how these modifications propagate to the final observables after the inclusion of hadronic rescattering and resonance decays.

\subsection{Effects on particle sampling}
\label{subsect:no-decay}
We now present results for final-state observables typically evaluated in hybrid simulations of heavy-ion collisions. To isolate the effect of the novel RTA particlization scheme, we first present the results obtained immediately following the particlization stage of the simulation chain, before hadronic decays are performed.
\begin{figure}[t!]
    \centering
    \includegraphics[trim={0.2cm 0.43cm 0.8cm 0.2cm}, clip, width=0.98\linewidth]{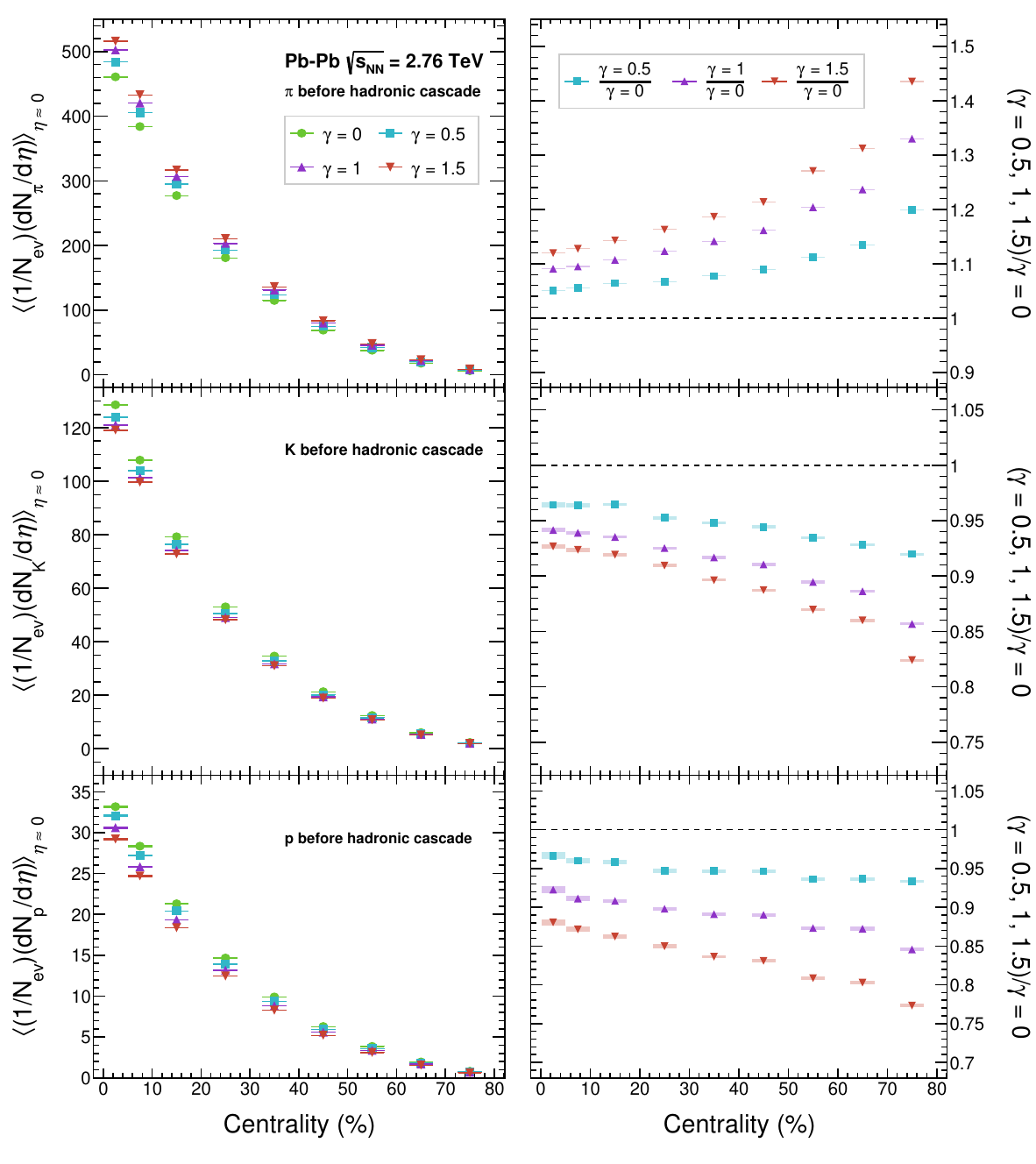}    
    \caption{Pion (top row), kaon (middle row) and proton (bottom row) multiplicity at mid-rapidity as a function of event centrality from hybrid simulations of Pb-Pb collisions at $\sqrt{s_\text{NN}}=2.76$ TeV (left column), for different values of $\gamma$, and its ratio to $\gamma = 0$ (right column), before hadronic decays.}
    \label{fig:pi-k-p-nd-PbPb}
\end{figure}

In Fig.~\ref{fig:pi-k-p-nd-PbPb}, we present the multiplicity of sampled pions, kaons, and protons (left column) at mid-rapidity for different values of the parameter $\gamma$ for Pb-Pb simulations at $\sqrt{s_\text{NN}}=2.76$ TeV. The relative differences with respect to the $\gamma = 0$ baseline are highlighted in the right column of Fig.~\ref{fig:pi-k-p-nd-PbPb}. In agreement with the computations for the species dependence of the multiplicity corrections presented in Fig.~\ref{fig:theory-deltaNs}, we note an increase in pion production and a decrease in both kaon and proton production as the value of $\gamma$ increases, as a result of the non-trivial competing effects shown in Fig.~\ref{fig:theory-deltaNs}.

\begin{figure}[t!]
    \centering
    \includegraphics[trim={0.1cm 0.05cm 0cm 0cm}, clip, width=0.98\linewidth]{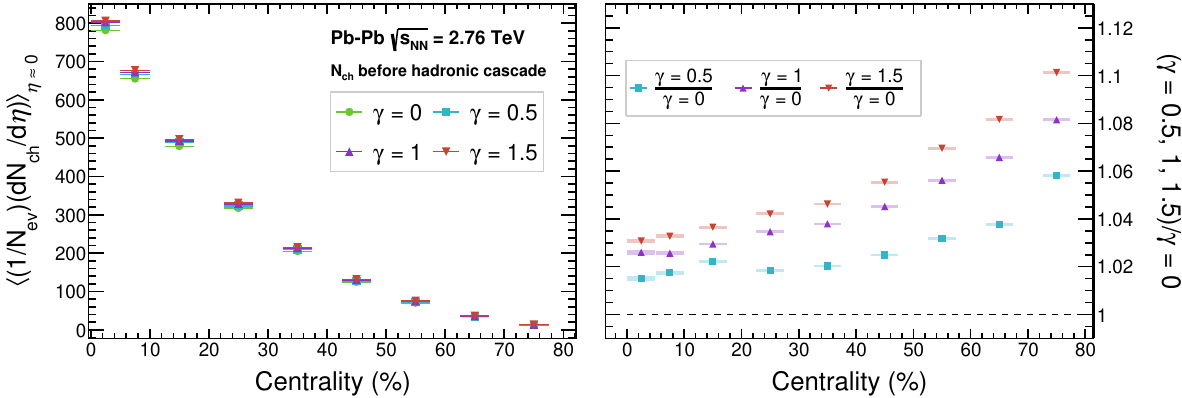}
    \caption{Charged-particle multiplicity at mid-rapidity as a function of event centrality from hybrid simulations of Pb-Pb collisions at $\sqrt{s_\text{NN}}=2.76$ TeV (left), for different values of $\gamma$, and its ratio to $\gamma = 0$ (right), before hadronic decays.}
    \label{fig:nch-nd-PbPb}
\end{figure}

The effect of the $\gamma$ parameter on the charged-particle multiplicity is shown in Fig.~\ref{fig:nch-nd-PbPb}. Consistent with the expected corrections presented in Fig.~\ref{fig:theory-deltaNs} (fourth panel) at $T \simeq 0.15$ GeV, the total multiplicity increases with $\gamma$. However, due to the competition among different hadron species discussed earlier, the net modification of the inclusive charged-particle multiplicity is significantly smaller than the corresponding changes observed for individually identified species. This cancellation is not accidental but a direct consequence of the species-dependent viscous corrections introduced by the generalized RTA.
\begin{figure}[t!]
    \centering
    \includegraphics[trim={0.2cm 0.05cm 0.6cm 0.2cm}, clip, width=0.99\linewidth]{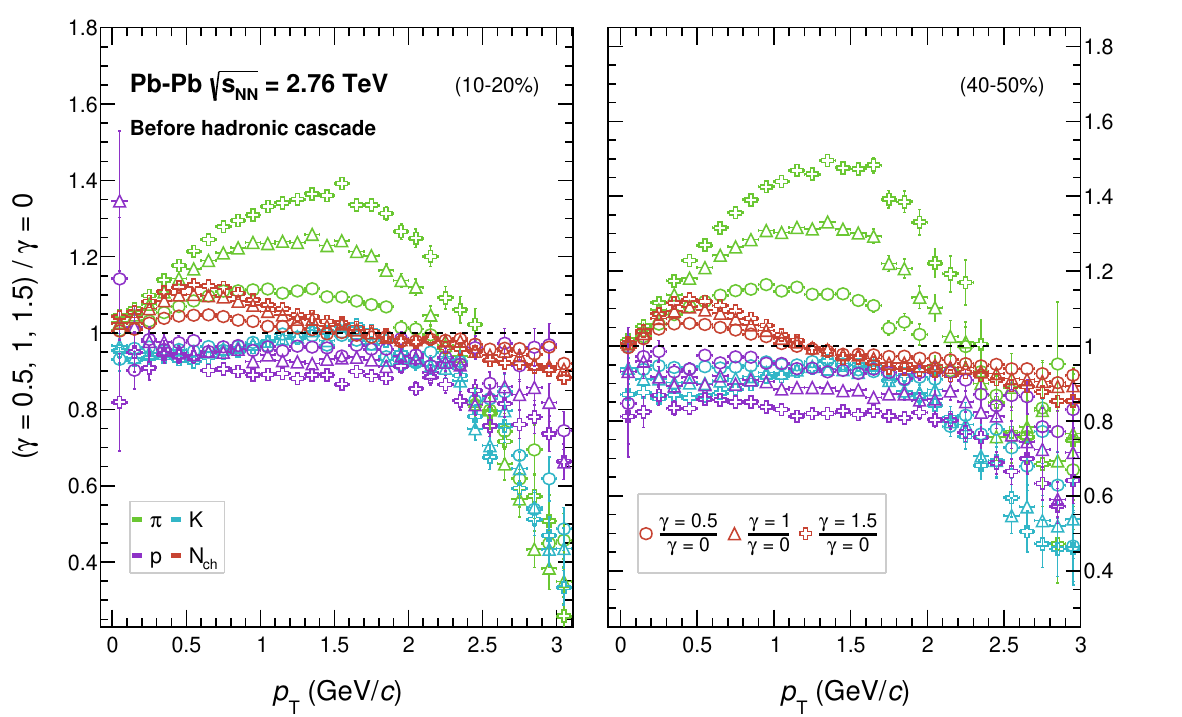}  
    \caption{$p_T$-spectra ratios for identified particles ($\pi, K, p$) and total charged particles, comparing different values of the $\gamma$ parameter ($\gamma = 0.5, 1.0, 1.5$) to the baseline value of the parameter ($\gamma = 0$). Results are obtained from hybrid simulations of Pb-Pb collisions at $\sqrt{s_{\text{NN}}} = 2.76$ TeV for 10-20\% (left) and 40-50\% (right) centrality classes. Markers denote $\gamma$ values, while colors distinguish particle species.}
    \label{fig:pTspectra-nd-PbPb}
\end{figure}
We also observe an effect of the novel RTA-based viscous corrections on particle transverse momentum spectra of identified particles. These effects on the transverse momentum distribution of identified particles are summarized in Fig.~\ref{fig:pTspectra-nd-PbPb} for Pb-Pb collision system. The results demonstrate that the enhancement or suppression of particle production is $p_T$ and species dependent, significantly altering the spectral shapes.

For pions (green curves), an increase in $\gamma$ leads to a noticeable enhancement at low $p_T$ ($p_T < 1.5$ GeV/c) and a progressive suppression of particles with $p_T > 2.5$ GeV/c. In contrast, protons (purple curves) show a different behavior. While the Pb-Pb system experiences a systematic suppression at low $p_T$. We have also studied p-Pb collisions at $\sqrt{s_\text{NN}}=5.02$ TeV. The corresponding results are presented in Appendix~\ref{sect:appendix_pPb}. This system also experiences a systematic suppression at low $p_T$, but it tends to recover or even exceed the baseline at higher $p_T$. The total charged-particle ratio (red curves) remains remarkably flat and close to unity across the $p_T$ range, especially in Pb-Pb collisions, indicating that the new RTA scheme redistributes momentum among species while maintaining the overall hydrodynamic flow profile.

Notably, despite the shorter lifetime and larger relative viscous corrections characteristic of p–Pb collisions, the species-dependent effects induced by the generalized RTA remain visible; see Appendix~\ref{sect:appendix_pPb}. As observed in Fig.~\ref{fig:pi-k-p-nd-PbPb}, the suppression of kaons and protons relative to pions (i.e., a decrease in $K/\pi$ and $p/\pi$ ratios) becomes more pronounced as $\gamma$ increases. This noted sensitivity of identified particle production to this new parameter $\gamma$ due to species mass dependence of the multiplicity corrections opens the door for improving the description of relative particle yields within hybrid models.

\subsection{Effects of hadronic decays}
\label{subsect:after-decays}

In realistic heavy-ion collision scenarios, a significant fraction of the final-state stable hadrons originates from the decays of heavier resonances. These decays can kinematically change momentum distributions and modify particle yields. To assess the phenomenological relevance of the nRTA signatures in hybrid simulations, it is important to determine if they persist after the system undergoes the hadronic cascade stage of the evolution. We now present the results on final state observables, after hadronic decays are performed on the sampled particles. In our chain, those were performed in UrQMD, together with hadronic rescatterings. 
\begin{figure}[t!]
    \centering
    \includegraphics[trim={0.2cm 0.43cm 0cm 0.2cm}, clip, width=0.98\linewidth]{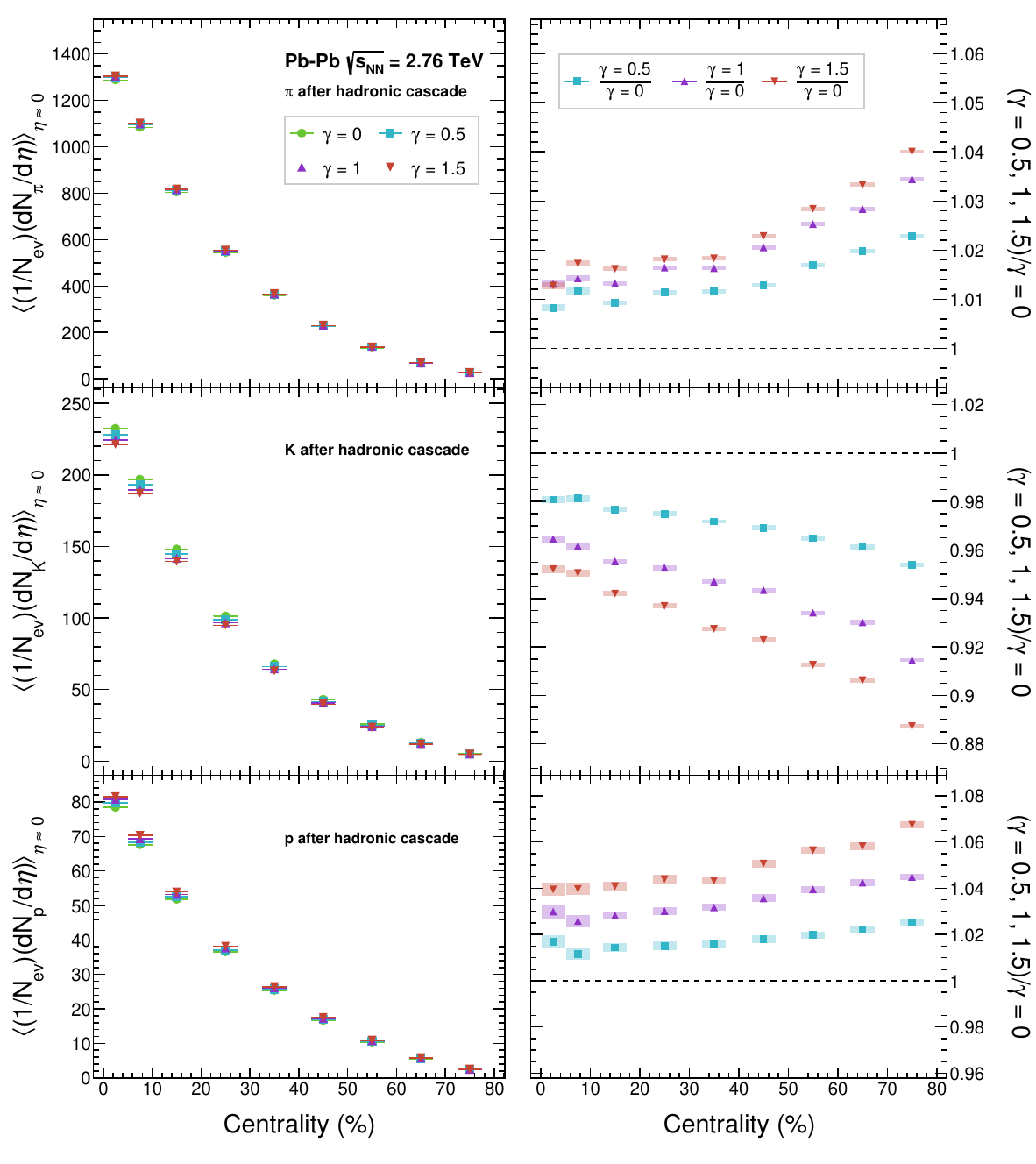}    
    \caption{Pion (top row), kaon (middle row) and proton (bottom row) multiplicity at mid-rapidity as a function of event centrality from hybrid simulations of Pb-Pb collisions at $\sqrt{s_\text{NN}}=2.76$ TeV (left column), for different values of $\gamma$, and its ratio to $\gamma = 0$ (right column), after hadronic decays.}
    \label{fig:pi-k-p-usual-PbPb}
\end{figure}

The final multiplicity of identified particles is presented in Fig.~\ref{fig:pi-k-p-usual-PbPb}, with the corresponding ratios relative to the $\gamma=0$ baseline. We observe that the multiplicity trends for pions and kaons are still present, albeit with a smaller magnitude. Notably, the trend is reversed for protons, possibly due to increased decays into protons and baryon-antibaryon annihilation during UrQMD. This effect can be observed across different centrality classes. 
\begin{figure}[t!]
    
    \includegraphics[trim={0.1cm 0.43cm 0cm 0cm}, clip, width=0.98\linewidth]{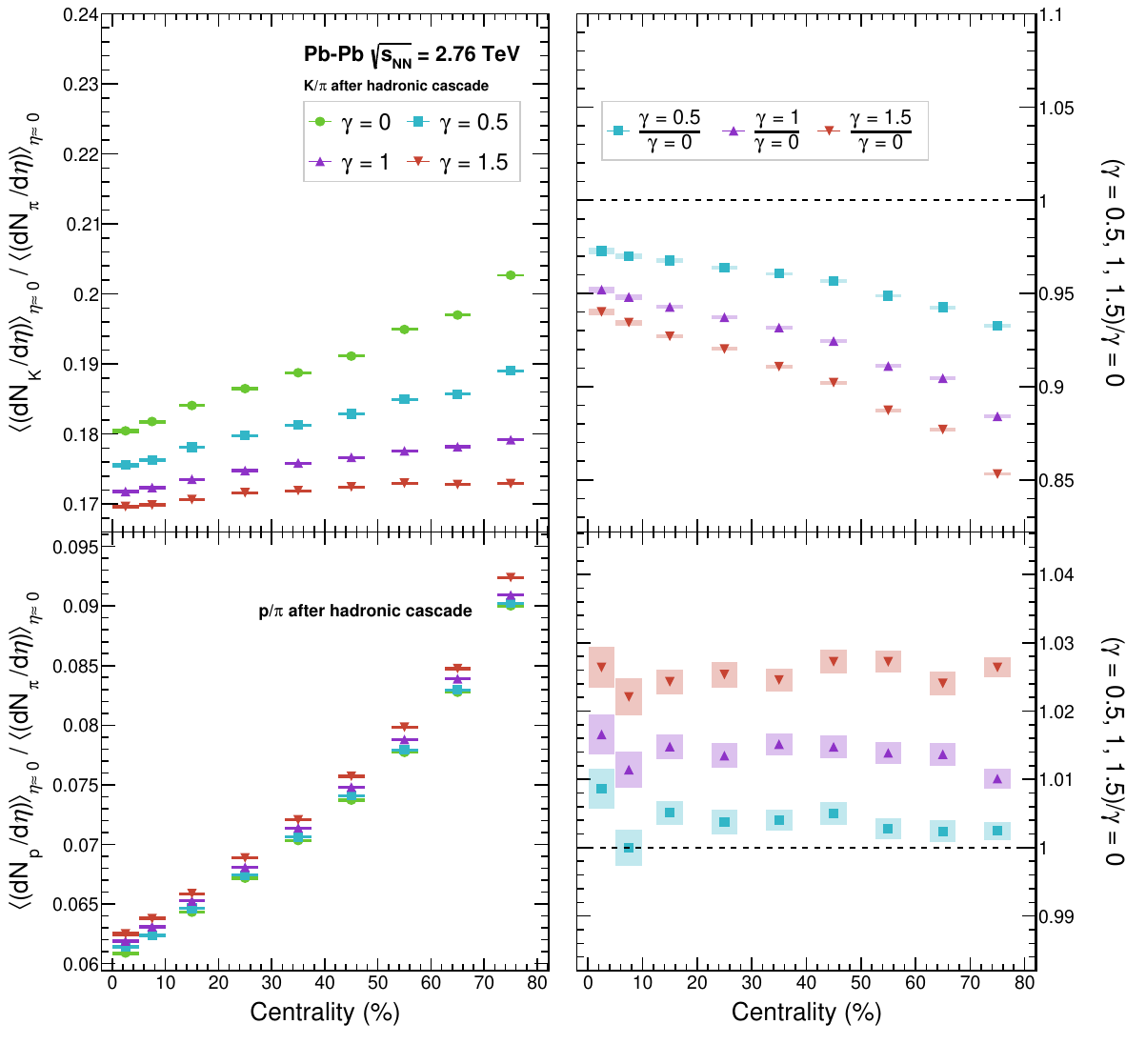}  
    \caption{Relative yield of kaons (top) and protons (bottom) as a function of event centrality from hybrid simulations of Pb-Pb collisions at $\sqrt{s_\text{NN}}=2.76$ TeV, after hadronic decays and with different values of the phenomenological parameter $\gamma$.}
    \label{fig:ratio-dNspec-usual-PbPb}
\end{figure}

\begin{figure}[t!]
    \centering
    \includegraphics[trim={0.1cm 0.05cm 0cm 0cm}, clip, width=0.98\linewidth]{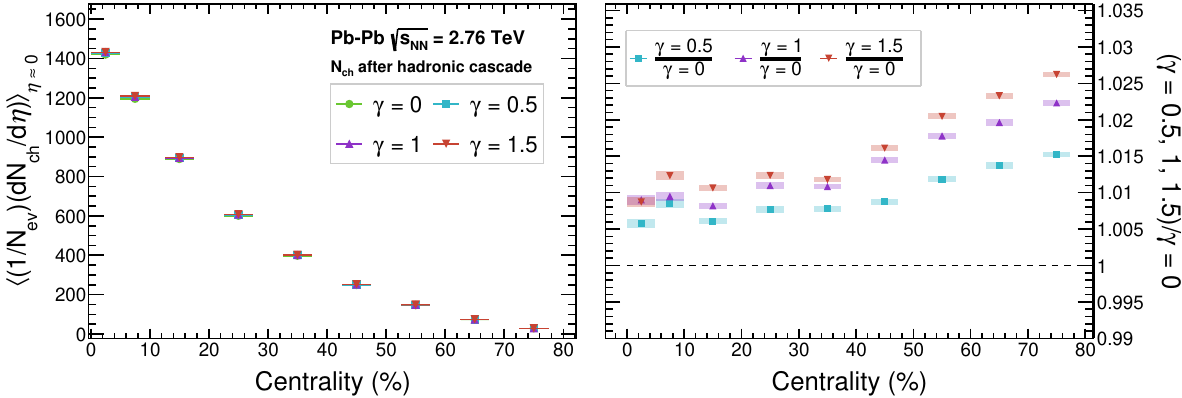}
    \caption{Charged-particle multiplicity at mid-rapidity as a function of event centrality from hybrid simulations of Pb-Pb collisions at $\sqrt{s_\text{NN}}=2.76$ TeV (left), for different values of $\gamma$, and its ratio to $\gamma = 0$ (right), after hadronic decays.}
    \label{fig:nch-usual-PbPb}
\end{figure}

The total multiplicity of charged particles after the decays is shown in Fig. ~\ref{fig:nch-usual-PbPb}. The overall increase with $\gamma$ remains visible, although the effect is diluted. This is due to the species dependence introduced into these observables by the new RTA formulation. The viscous corrections do not alter the total charged particle multiplicity significantly, but rather the relative yield of the different species.

\begin{figure}[t!]
    \centering
    \includegraphics[trim={0cm 0.cm 0cm 0cm}, clip, width=0.99\linewidth]{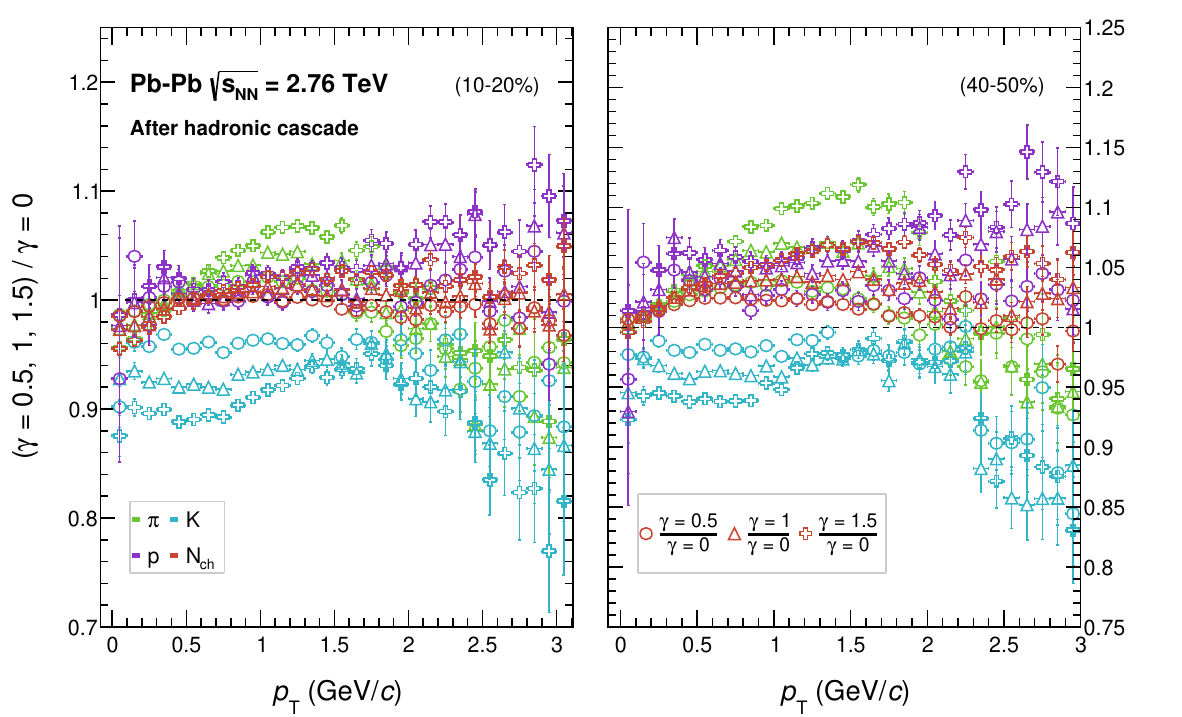}  
    \caption{$p_T$-spectra ratios for identified particles ($\pi, K, p$) and total charged particles, comparing different values of the $\gamma$ parameter ($\gamma = 0.5, 1.0, 1.5$) to the baseline value of the parameter ($\gamma = 0$). Results are obtained from hybrid simulations of Pb-Pb collisions at $\sqrt{s_{\text{NN}}} = 2.76$ TeV for 10-20\% (left) and 40-50\% (right) centrality classes. Markers denote $\gamma$ values, while colors distinguish particle species.}
    \label{fig:pTspectra-usual-PbPb}
\end{figure}

The differential transverse momentum spectra, including decays, are presented in Fig.~\ref{fig:pTspectra-usual-PbPb}. For charged particles, pions, and kaons, we observe the same trends reported in the spectra before hadronic decays, with a diminished magnitude once again. For protons, however, the spectra ratio rises above unity. 

\newpage

A systematic summary of analyzed observables is provided in Table~\ref{tab:summary}. 

\begin{table*}[t!]
\caption{Summary of effects in observables of nRTA particlization for Pb-Pb and p-Pb collisions.}
\label{tab:summary}
\begin{ruledtabular}
\begin{tabular}{lll}
    \multicolumn{1}{c}{\textbf{Observable}} & \multicolumn{1}{c}{\textbf{Before hadronic cascade}} & \multicolumn{1}{c}{\textbf{After hadronic cascade}} \\ \hline
    
    \begin{minipage}[t]{0.15\linewidth} \vspace{2pt} \raggedright Identified multiplicity \vspace{2pt} \end{minipage} & 
    \begin{minipage}[t]{0.4\linewidth} \vspace{2pt} \sloppy $\pi$ enhancement, $p$ and $K$ depletion with nRTA-$\gamma$; larger effects at more peripheral collisions. Fig.~\ref{fig:pi-k-p-nd-PbPb} (Fig.~\ref{fig:pi-k-p-nd-pPb}, p-Pb). \vspace{2pt} \end{minipage} & 
    \begin{minipage}[t]{0.4\linewidth} \vspace{2pt} \sloppy Weaker $\pi$ depletion and $p$ and $K$ enhancement with nRTA-$\gamma$; larger effects at more peripheral collisions. Figs.~\ref{fig:pi-k-p-usual-PbPb}, \ref{fig:ratio-dNspec-usual-PbPb} (Figs.~\ref{fig:pi-k-p-usual-pPb}, \ref{fig:ratio-dNspec-usual-pPb}, p-Pb). \vspace{2pt} \end{minipage} \\ \hline
    
    \begin{minipage}[t]{0.15\linewidth} \vspace{2pt} \raggedright Charged multiplicity \vspace{2pt} \end{minipage} & 
    \begin{minipage}[t]{0.4\linewidth} \vspace{2pt} \sloppy Enhancement with nRTA-$\gamma$, larger effects at more peripheral collisions for Pb-Pb, weak centrality dependence for p-Pb. Fig.~\ref{fig:nch-usual-PbPb} (Fig.~\ref{fig:nch-usual-pPb}, p-Pb). \vspace{2pt} \end{minipage} & 
    \begin{minipage}[t]{0.4\linewidth} \vspace{2pt} \sloppy Weaker enhancement with nRTA-$\gamma$, slightly larger effects at more peripheral collisions for Pb-Pb, slightly smaller effects for p-Pb. Fig.~\ref{fig:nch-usual-PbPb} (Fig.~\ref{fig:nch-usual-pPb}, p-Pb). \vspace{2pt} \end{minipage} \\ \hline
    
    \begin{minipage}[t]{0.15\linewidth} \vspace{2pt} \raggedright $p_{T}$ spectra \vspace{2pt} \end{minipage} & 
    \begin{minipage}[t]{0.4\linewidth} \vspace{2pt} \sloppy Enhancement of charged particles with $0.25 < p_T < 0.75$ with nRTA $\gamma$ in central collisions. Slight enhancement (depletion) of soft pions (kaons). Fig.~\ref{fig:pTspectra-nd-PbPb} (Fig.~\ref{fig:pTspectra-nd-pPb}, p-Pb). \vspace{2pt} \end{minipage} & 
    \begin{minipage}[t]{0.4\linewidth} \vspace{2pt} \sloppy Weaker enhancement of charged particles with $0.25 < p_T < 0.75$ with nRTA $\gamma$ in central collisions. Slight enhancement (depletion) of soft pions (kaons); flatter spectra. Fig.~\ref{fig:pTspectra-usual-PbPb} (Fig.~\ref{fig:pTspectra-usual-pPb}, p-Pb). \vspace{2pt} \end{minipage} \\
\end{tabular}
\end{ruledtabular}
\end{table*}

\section{Conclusion}
\label{sect:conclusions}

In this work, we applied a recently developed generalized relaxation time approximation (RTA) for multi-species relativistic gas \cite{Rocha:2025rkl} to realistic hybrid simulations of $p-\mathrm{Pb}$ and $\mathrm{Pb}-\mathrm{Pb}$ collisions. This novel formulation addresses known inconsistencies in the traditional Anderson-Witting RTA by rigorously ensuring local conservation of energy and momentum, even with momentum-dependent relaxation times $\tau_i(p)$. This rigorous kinetic constraint results in first-order viscous corrections ($\delta f_i$) that are explicitly species- and mass-dependent.

We demonstrated that these modified $\delta f_i$ corrections lead to significant and non-trivial changes in identified hadron production at the level of Cooper-Frye sampling. Specifically, both the yields and transverse-momentum ($p_T$) spectra of light hadrons ($\pi, K, p$) are affected, with the magnitude and sign of the correction term exhibiting an oscillatory dependence on particle mass. As a result, inclusive charged-particle observables show smaller modifications.

The most important phenomenological consequence is the resulting modification of relative particle yields, such as the $K/\pi$ and $p/\pi$ ratios. Since these ratios are central to characterizing the chemical composition of the medium and are key observables for the baryon-to-meson enhancement (baryon anomaly), our results strongly indicate that species-dependent viscous corrections at particlization must be considered in precision analyses seeking to disentangle kinetic freeze-out from hadronization mechanisms like coalescence. While the magnitude of the observed effects is reduced by the inclusion of the hadronic cascade, the imprint of the generalized RTA remains visible in the final-state observables. 

Regarding collective dynamics, while current results for large systems show no immediate large-scale distortions, the full impact of the generalized RTA on anisotropic flow coefficients, such as $v_2$, is an open question \cite{Rocha:2021zcw}. Preliminary tests suggest that potential effects may be more pronounced in small, high-gradient systems like p-Pb, where the sensitivity to the microscopic phase-space structure at particlization is significantly enhanced \cite{Weller:2017tsr}. A comprehensive systematic study of these flow observables, requiring larger statistics to resolve subtle differences in $v_n$, will be the subject of forthcoming work.

Beyond the phenomenology presented, our results open several compelling directions for future work. First, the species- and momentum-sensitive $\delta f$ introduces a new kinetic source of uncertainty; thus, our framework is essential to incorporate the associated parameters into Bayesian calibration studies to fully quantify the particlization uncertainty and its correlations with extracted QCD transport coefficients. Second, the mass-dependent $\delta f$ correction offers a kinetic mechanism that directly impacts the $p/\pi$ ratio and must be quantitatively compared against enhancement mechanisms in hadron coalescence models to rigorously disentangle the effects of final-state kinetic sampling from genuine hadronization effects. Finally, given that observables like electromagnetic radiation, heavy-flavor hadronization, and conserved-charge fluctuations can be  sensitive to the microscopic phase-space distribution $\delta f$ at the hydrodynamic/hadronic interface, our framework provides a robust, low-bias method to revisit these high-precision probes. In future studies, it will also be interesting to assess finite baryon chemical potential effects.

\section*{Acknowledgments}

I. A. is supported by Fundação de Amparo à Pesquisa e Inovação do Estado de Santa Catarina (FAPESC) grant 733/2024. I.A., T. N.dS. and G.S.D. are supported by CNPq through the INCT-FNA grant 408419/2024-5. T.N.dS. was supported by the Universal Grant 409029/2021-1. G.~S.~R. was funded by Vanderbilt University and was also funded by in part by the U.S. Department of Energy, Office of Science under Award Number DE-SC-0024347.  M.L.~was supported by the S\~{a}o Paulo Research Foundation (FAPESP) under projects 2018/24720-6, 2020/04867-2, and 2023/13749-1.
C.S. was supported by the Department of Energy, Office of Science, under DOE Award No.~DE-SC0021969. C.S. acknowledges a DOE Office of Science Early Career Award.

\appendix

\section{Effects on particle sampling and hadronic decays in p-Pb collision system}
\label{sect:appendix_pPb}

In this appendix, we complement the discussion presented in Sec.~\ref{sect:results} by showing the results for p-Pb collisions at $\sqrt{s_\text{NN}} = 5.02$ TeV. Small collision systems are characterized by steeper pressure gradients and shorter lifetimes compared to Pb-Pb system, which typically results in larger viscous corrections relative to the equilibrium background. Thus, investigating the impact of the nRTA framework in p-Pb is important to test the stability and limits of the species-dependent $\delta f$ corrections.

\begin{figure}[t!]
    \centering
    \includegraphics[trim={0.3cm 0.43cm 0cm 0cm}, clip, width=0.98\linewidth]{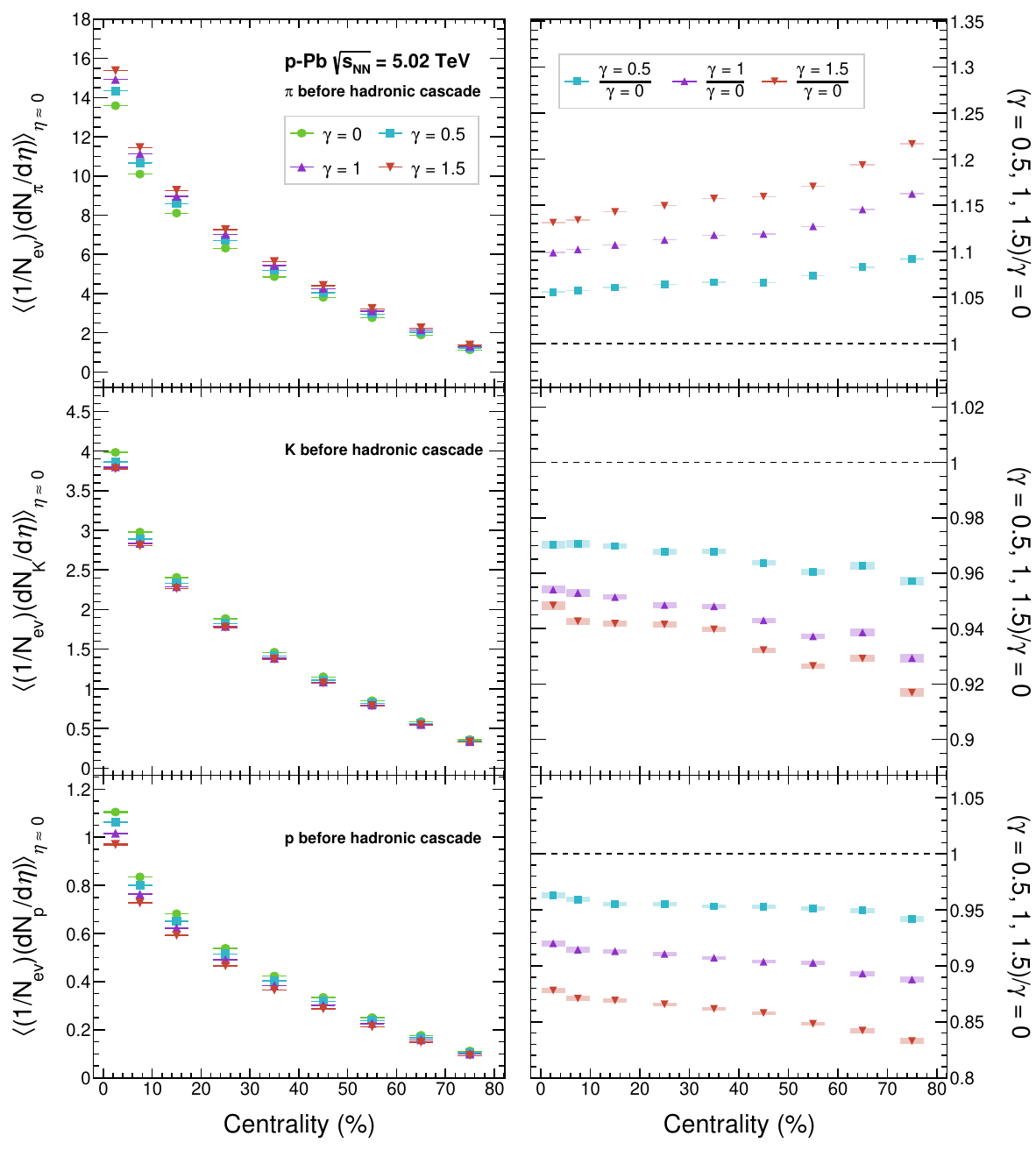}    
    \caption{Pion (top row), kaon (middle row) and proton (bottom row) multiplicity at mid-rapidity as a function of event centrality from hybrid simulations of p-Pb collisions at $\sqrt{s_\text{NN}} = 5.02$ TeV (left column), for different values of $\gamma$, and its ratio to $\gamma = 0$ (right column), before hadronic decays.}
    \label{fig:pi-k-p-nd-pPb}
\end{figure}

\begin{figure}[t!]
    \centering
    \includegraphics[trim={0.1cm 0.05cm 0cm 0cm}, clip, width=0.98\linewidth]{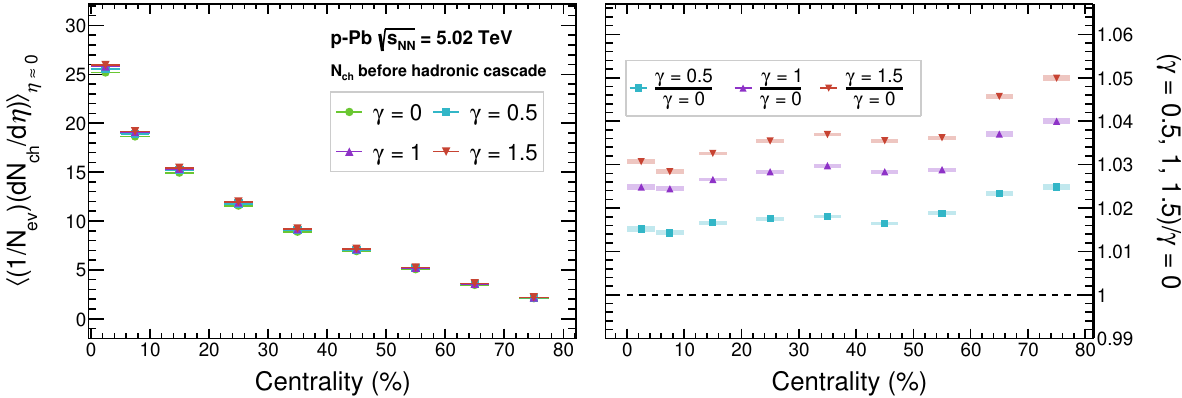}
    \caption{Charged-particle multiplicity at mid-rapidity as a function of event centrality from hybrid simulations of p-Pb collisions at $\sqrt{s_\text{NN}} = 5.02$ TeV (left), for different values of $\gamma$, and its ratio to $\gamma = 0$ (right), before hadronic decays.}
    \label{fig:nch-nd-pPb}
\end{figure}

Figures \ref{fig:pi-k-p-nd-pPb} and \ref{fig:nch-nd-pPb} show the identified and charged particle multiplicities at mid-rapidity before hadronic decays. It is observed that the hierarchical effect in the suppression/enhancement of yields according to the parameter $\gamma$ follows the same qualitative pattern as observed in Pb-Pb system, although the centrality dependence is less pronounced. This suggests that the species-dependent redistribution of yields is independent of the system size.

The transverse momentum spectra for the p-Pb system, shown in Fig.~\ref{fig:pTspectra-nd-PbPb} (before decays) and Fig.~\ref{fig:pTspectra-usual-pPb} (after decays), exhibit a similar crossing pattern in the ratios relative to $\gamma=0$. Interestingly, for protons in p-Pb, the recovery of the yield at higher $p_T$ is more visible than in the larger system. This behavior highlights that the choice of the relaxation time's energy dependence, i.e. $\gamma$, can modify the slope of the spectra in high-gradient environments.

Finally, the results after the hadronic decays with UrQMD, shown in Figs.~\ref{fig:pi-k-p-usual-pPb}, \ref{fig:ratio-dNspec-usual-pPb}, and \ref{fig:nch-usual-pPb}, confirm that the hadronic rescattering stage dilutes but does not erase the initial non-equilibrium signatures. The persistence of these effects in p-Pb system reinforces the importance of a consistent treatment of $\delta f$ when extracting transport properties from small system observables.
%\newpage
%
%\newpage
%
\begin{figure}[htb!]
    \centering
    \includegraphics[trim={0.2cm 0.05cm 0cm 0cm}, clip, width=0.99\linewidth]{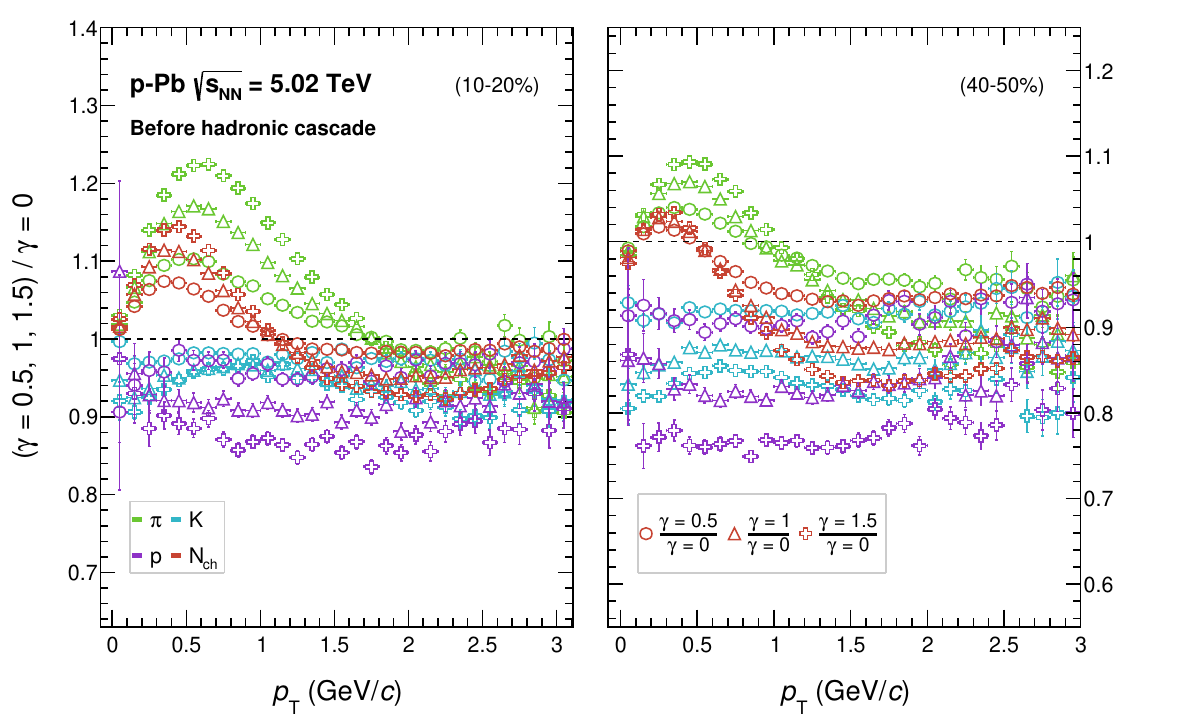}  
    \caption{$p_T$-spectra ratios for identified particles ($\pi, K, p$) and total charged particles, comparing different values of the $\gamma$ parameter ($\gamma = 0.5, 1.0, 1.5$) to the baseline value of the parameter ($\gamma = 0$). Results are obtained from hybrid simulations of p-Pb collisions at $\sqrt{s_{\text{NN}}} = 5.02$ TeV for 10-20\% (left) and 40-50\% (right) centrality classes. Markers denote $\gamma$ values, while colors distinguish species.
    }
    \label{fig:pTspectra-nd-pPb}
\end{figure}
%\newpage
%
\begin{figure}[t!]
    \centering
    \includegraphics[trim={0.2cm 0.05cm 0cm 0cm}, clip, width=0.99\linewidth]{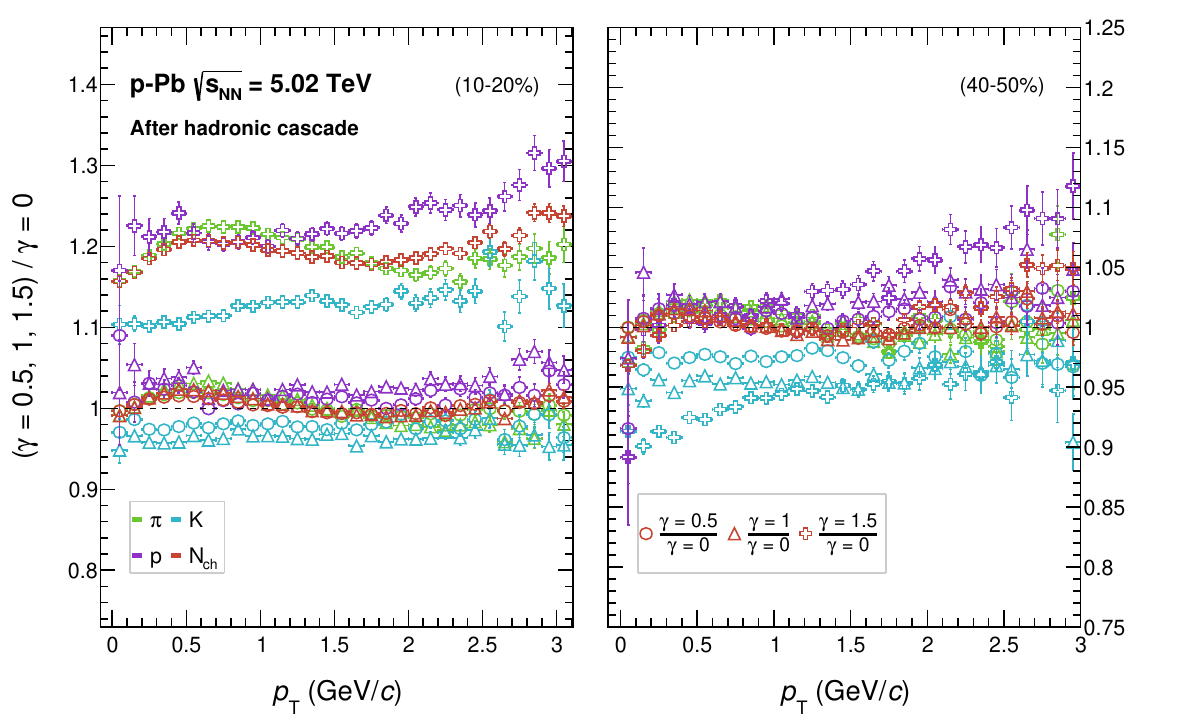}  
    \caption{$p_T$-spectra ratios for identified particles ($\pi, K, p$) and total charged particles, comparing different values of the $\gamma$ parameter ($\gamma = 0.5, 1.0, 1.5$) to the baseline value of the parameter ($\gamma = 0$). Results are obtained from hybrid simulations of p-Pb collisions at $\sqrt{s_{\text{NN}}} = 5.02$ TeV for 10-20\% (left) and 40-50\% (right) centrality classes. Markers denote $\gamma$ values, while colors distinguish particle species.}
    \label{fig:pTspectra-usual-pPb}
\end{figure}
%
%\newpage
%
\begin{figure}[t!]
    \centering
    \includegraphics[trim={0.3cm 0.43cm 0cm 0cm}, clip, width=0.98\linewidth]{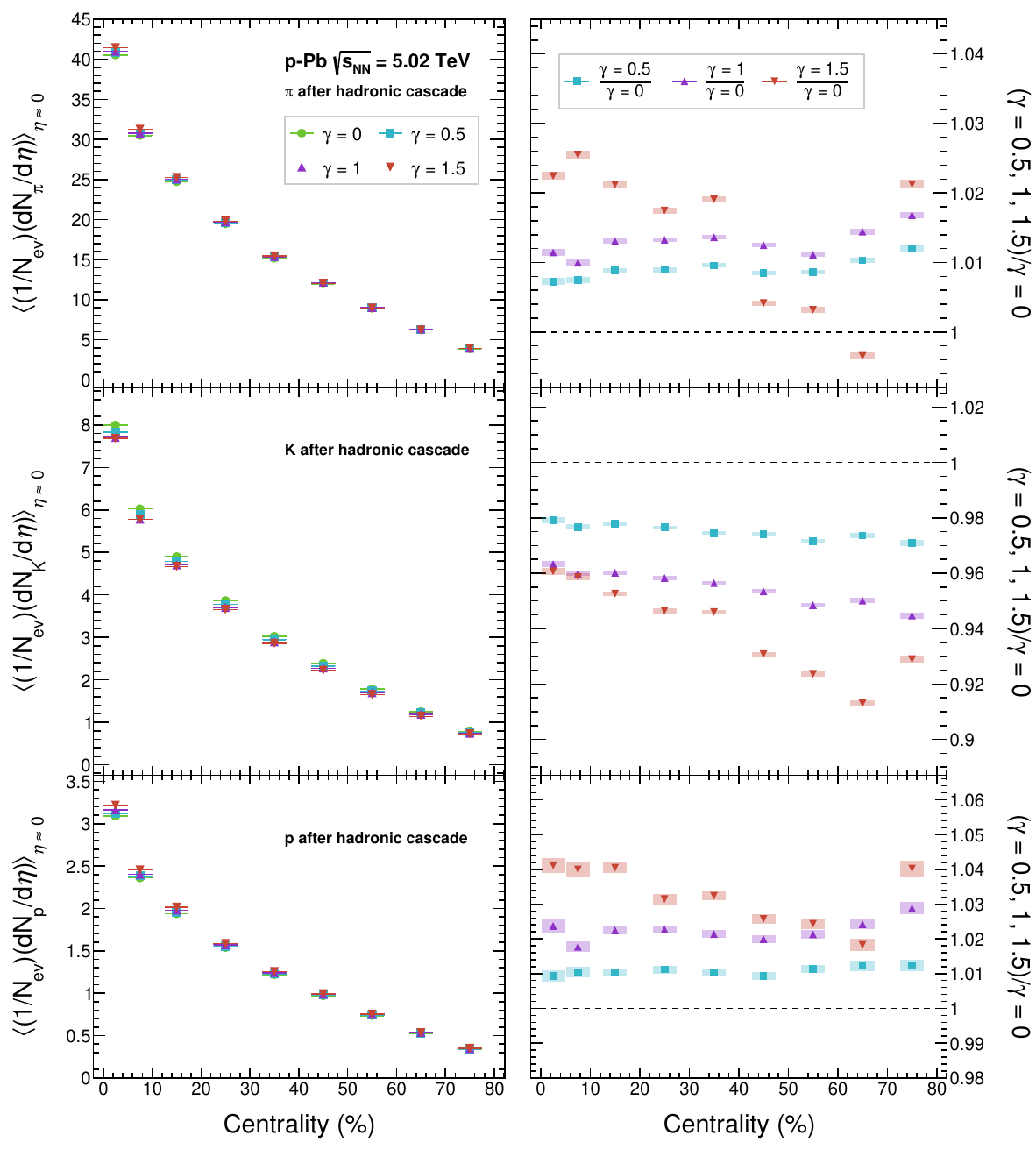}    
    \caption{Pion (top row), kaon (middle row) and proton (bottom row) multiplicity at mid-rapidity as a function of event centrality from hybrid simulations of p-Pb collisions at $\sqrt{s_\text{NN}} = 5.02$ TeV (left column), for different values of $\gamma$, and its ratio to $\gamma = 0$ (right column), after hadronic decays.}
    \label{fig:pi-k-p-usual-pPb}
\end{figure}
%
%\newpage
%
\begin{figure}[t!]
    \includegraphics[trim={0.1cm 0.43cm 0cm 0cm}, clip, width=0.98\linewidth]{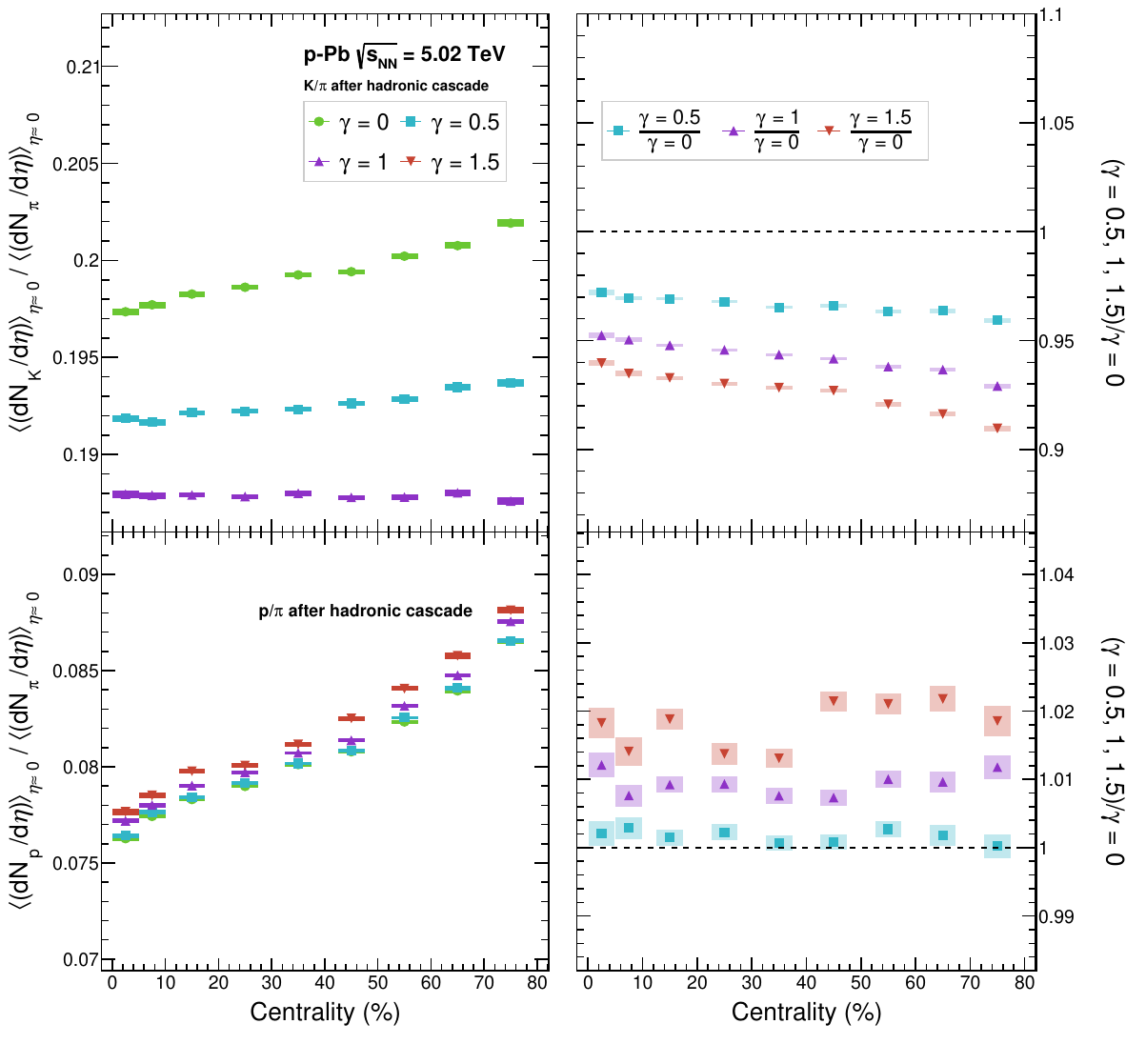}  
    \caption{Relative yield of kaons (top) and protons (bottom) as a function of event centrality from hybrid simulations of p-Pb collisions at $\sqrt{s_\text{NN}} = 5.02$ TeV, after hadronic decays and with different values of the phenomenological parameter $\gamma$}
    \label{fig:ratio-dNspec-usual-pPb}
\end{figure}
%
%\newpage
\begin{figure}[t!]
    \centering
    \includegraphics[trim={0.1cm 0.05cm 0cm 0cm}, clip, width=0.98\linewidth]{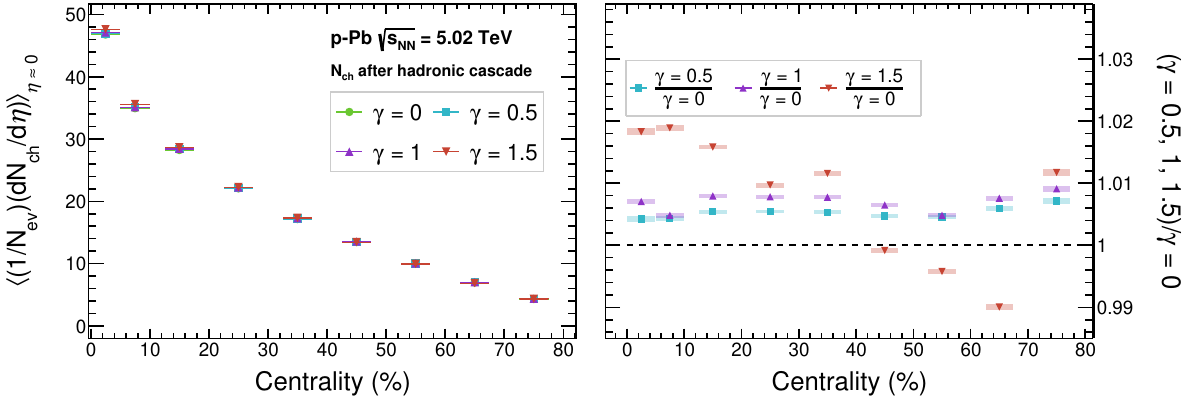}
    \caption{Charged-particle multiplicity at mid-rapidity as a function of event centrality from hybrid simulations of p-Pb collisions at $\sqrt{s_\text{NN}} = 5.02$ TeV (left), for different values of $\gamma$, and its ratio to $\gamma = 0$ (right), after hadronic decays.}
    \label{fig:nch-usual-pPb}
\end{figure}
%
%\newpage

\FloatBarrier
\bibliography{ref.bib}% Produces the bibliography via BibTeX.

\begin{thebibliography}{60}%
\makeatletter
\providecommand \@ifxundefined [1]{%
 \@ifx{#1\undefined}
}%
\providecommand \@ifnum [1]{%
 \ifnum #1\expandafter \@firstoftwo
 \else \expandafter \@secondoftwo
 \fi
}%
\providecommand \@ifx [1]{%
 \ifx #1\expandafter \@firstoftwo
 \else \expandafter \@secondoftwo
 \fi
}%
\providecommand \natexlab [1]{#1}%
\providecommand \enquote  [1]{``#1''}%
\providecommand \bibnamefont  [1]{#1}%
\providecommand \bibfnamefont [1]{#1}%
\providecommand \citenamefont [1]{#1}%
\providecommand \href@noop [0]{\@secondoftwo}%
\providecommand \href [0]{\begingroup \@sanitize@url \@href}%
\providecommand \@href[1]{\@@startlink{#1}\@@href}%
\providecommand \@@href[1]{\endgroup#1\@@endlink}%
\providecommand \@sanitize@url [0]{\catcode `\\12\catcode `\$12\catcode
  `\&12\catcode `\#12\catcode `\^12\catcode `\_12\catcode `\%12\relax}%
\providecommand \@@startlink[1]{}%
\providecommand \@@endlink[0]{}%
\providecommand \url  [0]{\begingroup\@sanitize@url \@url }%
\providecommand \@url [1]{\endgroup\@href {#1}{\urlprefix }}%
\providecommand \urlprefix  [0]{URL }%
\providecommand \Eprint [0]{\href }%
\providecommand \doibase [0]{https://doi.org/}%
\providecommand \selectlanguage [0]{\@gobble}%
\providecommand \bibinfo  [0]{\@secondoftwo}%
\providecommand \bibfield  [0]{\@secondoftwo}%
\providecommand \translation [1]{[#1]}%
\providecommand \BibitemOpen [0]{}%
\providecommand \bibitemStop [0]{}%
\providecommand \bibitemNoStop [0]{.\EOS\space}%
\providecommand \EOS [0]{\spacefactor3000\relax}%
\providecommand \BibitemShut  [1]{\csname bibitem#1\endcsname}%
\let\auto@bib@innerbib\@empty
%</preamble>
\bibitem [{\citenamefont {Gross}\ and\ \citenamefont
  {Wilczek}(1973)}]{Gross:1973id}%
  \BibitemOpen
  \bibfield  {author} {\bibinfo {author} {\bibfnamefont {D.~J.}\ \bibnamefont
  {Gross}}\ and\ \bibinfo {author} {\bibfnamefont {F.}~\bibnamefont
  {Wilczek}},\ }\bibfield  {title} {\bibinfo {title} {{Ultraviolet Behavior of
  Nonabelian Gauge Theories}},\ }\href
  {https://doi.org/10.1103/PhysRevLett.30.1343} {\bibfield  {journal} {\bibinfo
   {journal} {Phys. Rev. Lett.}\ }\textbf {\bibinfo {volume} {30}},\ \bibinfo
  {pages} {1343} (\bibinfo {year} {1973})}\BibitemShut {NoStop}%
\bibitem [{\citenamefont {Politzer}(1973)}]{Politzer:1973fx}%
  \BibitemOpen
  \bibfield  {author} {\bibinfo {author} {\bibfnamefont {H.}~\bibnamefont
  {Politzer}},\ }\bibfield  {title} {\bibinfo {title} {{Reliable Perturbative
  Results for Strong Interactions?}},\ }\href
  {https://doi.org/10.1103/PhysRevLett.30.1346} {\bibfield  {journal} {\bibinfo
   {journal} {Phys. Rev. Lett.}\ }\textbf {\bibinfo {volume} {30}},\ \bibinfo
  {pages} {1346} (\bibinfo {year} {1973})}\BibitemShut {NoStop}%
\bibitem [{\citenamefont {Shuryak}(1978)}]{Shuryak:1978ij}%
  \BibitemOpen
  \bibfield  {author} {\bibinfo {author} {\bibfnamefont {E.~V.}\ \bibnamefont
  {Shuryak}},\ }\bibfield  {title} {\bibinfo {title} {{Quark-Gluon Plasma and
  Hadronic Production of Leptons, Photons and Psions}},\ }\href
  {https://doi.org/10.1016/0370-2693(78)90370-2} {\bibfield  {journal}
  {\bibinfo  {journal} {Sov. J. Nucl. Phys.}\ }\textbf {\bibinfo {volume}
  {28}},\ \bibinfo {pages} {408} (\bibinfo {year} {1978})}\BibitemShut
  {NoStop}%
\bibitem [{\citenamefont {de~Forcrand}(2009)}]{deForcrand:2009zkb}%
  \BibitemOpen
  \bibfield  {author} {\bibinfo {author} {\bibfnamefont {P.}~\bibnamefont
  {de~Forcrand}},\ }\bibfield  {title} {\bibinfo {title} {{Simulating QCD at
  finite density}},\ }\href {https://doi.org/10.22323/1.091.0010} {\bibfield
  {journal} {\bibinfo  {journal} {PoS}\ }\textbf {\bibinfo {volume}
  {LAT2009}},\ \bibinfo {pages} {010} (\bibinfo {year} {2009})},\ \Eprint
  {https://arxiv.org/abs/1005.0539} {arXiv:1005.0539 [hep-lat]} \BibitemShut
  {NoStop}%
\bibitem [{\citenamefont {Heinz}\ and\ \citenamefont
  {Jacob}(2000)}]{Heinz:2000bk}%
  \BibitemOpen
  \bibfield  {author} {\bibinfo {author} {\bibfnamefont {U.~W.}\ \bibnamefont
  {Heinz}}\ and\ \bibinfo {author} {\bibfnamefont {M.}~\bibnamefont {Jacob}},\
  }\bibfield  {title} {\bibinfo {title} {{Evidence for a new state of matter:
  An Assessment of the results from the CERN lead beam program}},\ }\href@noop
  {} {\bibfield  {journal} {\bibinfo  {journal} {arXiv}\ } (\bibinfo {year}
  {2000})},\ \Eprint {https://arxiv.org/abs/nucl-th/0002042}
  {arXiv:nucl-th/0002042} \BibitemShut {NoStop}%
\bibitem [{\citenamefont {Arsene}\ \emph {et~al.}(2005)\citenamefont {Arsene}
  \emph {et~al.}}]{Arsene:2004fa}%
  \BibitemOpen
  \bibfield  {author} {\bibinfo {author} {\bibfnamefont {I.}~\bibnamefont
  {Arsene}} \emph {et~al.} (\bibinfo {collaboration} {BRAHMS}),\ }\bibfield
  {title} {\bibinfo {title} {{Quark gluon plasma and color glass condensate at
  RHIC? The Perspective from the BRAHMS experiment}},\ }\href
  {https://doi.org/10.1016/j.nuclphysa.2005.02.130} {\bibfield  {journal}
  {\bibinfo  {journal} {Nucl. Phys. A}\ }\textbf {\bibinfo {volume} {757}},\
  \bibinfo {pages} {1} (\bibinfo {year} {2005})},\ \Eprint
  {https://arxiv.org/abs/nucl-ex/0410020} {arXiv:nucl-ex/0410020} \BibitemShut
  {NoStop}%
\bibitem [{\citenamefont {Back}\ \emph {et~al.}(2005)\citenamefont {Back} \emph
  {et~al.}}]{Back:2004je}%
  \BibitemOpen
  \bibfield  {author} {\bibinfo {author} {\bibfnamefont {B.}~\bibnamefont
  {Back}} \emph {et~al.} (\bibinfo {collaboration} {PHOBOS}),\ }\bibfield
  {title} {\bibinfo {title} {{The PHOBOS perspective on discoveries at RHIC}},\
  }\href {https://doi.org/10.1016/j.nuclphysa.2005.03.084} {\bibfield
  {journal} {\bibinfo  {journal} {Nucl. Phys. A}\ }\textbf {\bibinfo {volume}
  {757}},\ \bibinfo {pages} {28} (\bibinfo {year} {2005})},\ \Eprint
  {https://arxiv.org/abs/nucl-ex/0410022} {arXiv:nucl-ex/0410022} \BibitemShut
  {NoStop}%
\bibitem [{\citenamefont {Adcox}\ \emph {et~al.}(2005)\citenamefont {Adcox}
  \emph {et~al.}}]{Adcox:2004mh}%
  \BibitemOpen
  \bibfield  {author} {\bibinfo {author} {\bibfnamefont {K.}~\bibnamefont
  {Adcox}} \emph {et~al.} (\bibinfo {collaboration} {PHENIX}),\ }\bibfield
  {title} {\bibinfo {title} {{Formation of dense partonic matter in
  relativistic nucleus-nucleus collisions at RHIC: Experimental evaluation by
  the PHENIX collaboration}},\ }\href
  {https://doi.org/10.1016/j.nuclphysa.2005.03.086} {\bibfield  {journal}
  {\bibinfo  {journal} {Nucl. Phys. A}\ }\textbf {\bibinfo {volume} {757}},\
  \bibinfo {pages} {184} (\bibinfo {year} {2005})},\ \Eprint
  {https://arxiv.org/abs/nucl-ex/0410003} {arXiv:nucl-ex/0410003} \BibitemShut
  {NoStop}%
\bibitem [{\citenamefont {Adams}\ \emph {et~al.}(2005)\citenamefont {Adams}
  \emph {et~al.}}]{Adams:2005dq}%
  \BibitemOpen
  \bibfield  {author} {\bibinfo {author} {\bibfnamefont {J.}~\bibnamefont
  {Adams}} \emph {et~al.} (\bibinfo {collaboration} {STAR}),\ }\bibfield
  {title} {\bibinfo {title} {{Experimental and theoretical challenges in the
  search for the quark gluon plasma: The STAR Collaboration's critical
  assessment of the evidence from RHIC collisions}},\ }\href
  {https://doi.org/10.1016/j.nuclphysa.2005.03.085} {\bibfield  {journal}
  {\bibinfo  {journal} {Nucl. Phys. A}\ }\textbf {\bibinfo {volume} {757}},\
  \bibinfo {pages} {102} (\bibinfo {year} {2005})},\ \Eprint
  {https://arxiv.org/abs/nucl-ex/0501009} {arXiv:nucl-ex/0501009} \BibitemShut
  {NoStop}%
\bibitem [{\citenamefont {Muller}\ \emph {et~al.}(2012)\citenamefont {Muller},
  \citenamefont {Schukraft},\ and\ \citenamefont {Wyslouch}}]{Muller:2012zq}%
  \BibitemOpen
  \bibfield  {author} {\bibinfo {author} {\bibfnamefont {B.}~\bibnamefont
  {Muller}}, \bibinfo {author} {\bibfnamefont {J.}~\bibnamefont {Schukraft}},\
  and\ \bibinfo {author} {\bibfnamefont {B.}~\bibnamefont {Wyslouch}},\
  }\bibfield  {title} {\bibinfo {title} {{First Results from Pb+Pb collisions
  at the LHC}},\ }\href {https://doi.org/10.1146/annurev-nucl-102711-094910}
  {\bibfield  {journal} {\bibinfo  {journal} {Ann. Rev. Nucl. Part. Sci.}\
  }\textbf {\bibinfo {volume} {62}},\ \bibinfo {pages} {361} (\bibinfo {year}
  {2012})},\ \Eprint {https://arxiv.org/abs/1202.3233} {arXiv:1202.3233
  [hep-ex]} \BibitemShut {NoStop}%
\bibitem [{\citenamefont {Abelev}\ \emph {et~al.}(2013)\citenamefont {Abelev}
  \emph {et~al.}}]{ALICE:2012eyl}%
  \BibitemOpen
  \bibfield  {author} {\bibinfo {author} {\bibfnamefont {B.}~\bibnamefont
  {Abelev}} \emph {et~al.} (\bibinfo {collaboration} {ALICE}),\ }\bibfield
  {title} {\bibinfo {title} {{Long-range angular correlations on the near and
  away side in $p$-Pb collisions at $\sqrt{s_{NN}}=5.02$ TeV}},\ }\href
  {https://doi.org/10.1016/j.physletb.2013.01.012} {\bibfield  {journal}
  {\bibinfo  {journal} {Phys. Lett. B}\ }\textbf {\bibinfo {volume} {719}},\
  \bibinfo {pages} {29} (\bibinfo {year} {2013})},\ \Eprint
  {https://arxiv.org/abs/1212.2001} {arXiv:1212.2001 [nucl-ex]} \BibitemShut
  {NoStop}%
\bibitem [{\citenamefont {Aad}\ \emph {et~al.}(2013)\citenamefont {Aad} \emph
  {et~al.}}]{ATLAS:2012cix}%
  \BibitemOpen
  \bibfield  {author} {\bibinfo {author} {\bibfnamefont {G.}~\bibnamefont
  {Aad}} \emph {et~al.} (\bibinfo {collaboration} {ATLAS}),\ }\bibfield
  {title} {\bibinfo {title} {{Observation of Associated Near-Side and Away-Side
  Long-Range Correlations in $\sqrt{s_{NN}}$=5.02 TeV Proton-Lead Collisions
  with the ATLAS Detector}},\ }\href
  {https://doi.org/10.1103/PhysRevLett.110.182302} {\bibfield  {journal}
  {\bibinfo  {journal} {Phys. Rev. Lett.}\ }\textbf {\bibinfo {volume} {110}},\
  \bibinfo {pages} {182302} (\bibinfo {year} {2013})},\ \Eprint
  {https://arxiv.org/abs/1212.5198} {arXiv:1212.5198 [hep-ex]} \BibitemShut
  {NoStop}%
\bibitem [{\citenamefont {Chatrchyan}\ \emph {et~al.}(2013)\citenamefont
  {Chatrchyan} \emph {et~al.}}]{CMS:2012qk}%
  \BibitemOpen
  \bibfield  {author} {\bibinfo {author} {\bibfnamefont {S.}~\bibnamefont
  {Chatrchyan}} \emph {et~al.} (\bibinfo {collaboration} {CMS}),\ }\bibfield
  {title} {\bibinfo {title} {{Observation of Long-Range Near-Side Angular
  Correlations in Proton-Lead Collisions at the LHC}},\ }\href
  {https://doi.org/10.1016/j.physletb.2012.11.025} {\bibfield  {journal}
  {\bibinfo  {journal} {Phys. Lett. B}\ }\textbf {\bibinfo {volume} {718}},\
  \bibinfo {pages} {795} (\bibinfo {year} {2013})},\ \Eprint
  {https://arxiv.org/abs/1210.5482} {arXiv:1210.5482 [nucl-ex]} \BibitemShut
  {NoStop}%
\bibitem [{\citenamefont {Adare}\ \emph {et~al.}(2013)\citenamefont {Adare}
  \emph {et~al.}}]{PHENIX:2013ktj}%
  \BibitemOpen
  \bibfield  {author} {\bibinfo {author} {\bibfnamefont {A.}~\bibnamefont
  {Adare}} \emph {et~al.} (\bibinfo {collaboration} {PHENIX}),\ }\bibfield
  {title} {\bibinfo {title} {{Quadrupole Anisotropy in Dihadron Azimuthal
  Correlations in Central $d+$Au Collisions at $\sqrt{s_{_{NN}}}$=200 GeV}},\
  }\href {https://doi.org/10.1103/PhysRevLett.111.212301} {\bibfield  {journal}
  {\bibinfo  {journal} {Phys. Rev. Lett.}\ }\textbf {\bibinfo {volume} {111}},\
  \bibinfo {pages} {212301} (\bibinfo {year} {2013})},\ \Eprint
  {https://arxiv.org/abs/1303.1794} {arXiv:1303.1794 [nucl-ex]} \BibitemShut
  {NoStop}%
\bibitem [{\citenamefont {Heinz}\ and\ \citenamefont
  {Snellings}(2013)}]{Heinz:2013th}%
  \BibitemOpen
  \bibfield  {author} {\bibinfo {author} {\bibfnamefont {U.}~\bibnamefont
  {Heinz}}\ and\ \bibinfo {author} {\bibfnamefont {R.}~\bibnamefont
  {Snellings}},\ }\bibfield  {title} {\bibinfo {title} {{Collective flow and
  viscosity in relativistic heavy-ion collisions}},\ }\href
  {https://doi.org/10.1146/annurev-nucl-102212-170540} {\bibfield  {journal}
  {\bibinfo  {journal} {Ann. Rev. Nucl. Part. Sci.}\ }\textbf {\bibinfo
  {volume} {63}},\ \bibinfo {pages} {123} (\bibinfo {year} {2013})},\ \Eprint
  {https://arxiv.org/abs/1301.2826} {arXiv:1301.2826 [nucl-th]} \BibitemShut
  {NoStop}%
\bibitem [{\citenamefont {Foka}\ and\ \citenamefont
  {Janik}(2016)}]{Foka:2016vta}%
  \BibitemOpen
  \bibfield  {author} {\bibinfo {author} {\bibfnamefont {P.}~\bibnamefont
  {Foka}}\ and\ \bibinfo {author} {\bibfnamefont {M.~g.~A.}\ \bibnamefont
  {Janik}},\ }\bibfield  {title} {\bibinfo {title} {{An overview of
  experimental results from ultra-relativistic heavy-ion collisions at the CERN
  LHC: Bulk properties and dynamical evolution}},\ }\href
  {https://doi.org/10.1016/j.revip.2016.11.002} {\bibfield  {journal} {\bibinfo
   {journal} {Rev. Phys.}\ }\textbf {\bibinfo {volume} {1}},\ \bibinfo {pages}
  {154} (\bibinfo {year} {2016})},\ \Eprint {https://arxiv.org/abs/1702.07233}
  {arXiv:1702.07233 [hep-ex]} \BibitemShut {NoStop}%
\bibitem [{\citenamefont {Loizides}(2016)}]{Loizides:2016tew}%
  \BibitemOpen
  \bibfield  {author} {\bibinfo {author} {\bibfnamefont {C.}~\bibnamefont
  {Loizides}},\ }\bibfield  {title} {\bibinfo {title} {{Experimental overview
  on small collision systems at the LHC}},\ }\href
  {https://doi.org/10.1016/j.nuclphysa.2016.04.022} {\bibfield  {journal}
  {\bibinfo  {journal} {Nucl. Phys. A}\ }\textbf {\bibinfo {volume} {956}},\
  \bibinfo {pages} {200} (\bibinfo {year} {2016})},\ \Eprint
  {https://arxiv.org/abs/1602.09138} {arXiv:1602.09138 [nucl-ex]} \BibitemShut
  {NoStop}%
\bibitem [{\citenamefont {Israel}(1976)}]{Israel:1976tn}%
  \BibitemOpen
  \bibfield  {author} {\bibinfo {author} {\bibfnamefont {W.}~\bibnamefont
  {Israel}},\ }\bibfield  {title} {\bibinfo {title} {{Nonstationary
  irreversible thermodynamics: A Causal relativistic theory}},\ }\href
  {https://doi.org/10.1016/0003-4916(76)90064-6} {\bibfield  {journal}
  {\bibinfo  {journal} {Annals Phys.}\ }\textbf {\bibinfo {volume} {100}},\
  \bibinfo {pages} {310} (\bibinfo {year} {1976})}\BibitemShut {NoStop}%
\bibitem [{\citenamefont {Israel}\ and\ \citenamefont
  {Stewart}(1979)}]{Israel:1979wp}%
  \BibitemOpen
  \bibfield  {author} {\bibinfo {author} {\bibfnamefont {W.}~\bibnamefont
  {Israel}}\ and\ \bibinfo {author} {\bibfnamefont {J.}~\bibnamefont
  {Stewart}},\ }\bibfield  {title} {\bibinfo {title} {{Transient relativistic
  thermodynamics and kinetic theory}},\ }\href
  {https://doi.org/10.1016/0003-4916(79)90130-1} {\bibfield  {journal}
  {\bibinfo  {journal} {Annals Phys.}\ }\textbf {\bibinfo {volume} {118}},\
  \bibinfo {pages} {341} (\bibinfo {year} {1979})}\BibitemShut {NoStop}%
\bibitem [{\citenamefont {Petersen}\ \emph {et~al.}(2008)\citenamefont
  {Petersen}, \citenamefont {Steinheimer}, \citenamefont {Burau}, \citenamefont
  {Bleicher},\ and\ \citenamefont {Stocker}}]{Petersen:2008dd}%
  \BibitemOpen
  \bibfield  {author} {\bibinfo {author} {\bibfnamefont {H.}~\bibnamefont
  {Petersen}}, \bibinfo {author} {\bibfnamefont {J.}~\bibnamefont
  {Steinheimer}}, \bibinfo {author} {\bibfnamefont {G.}~\bibnamefont {Burau}},
  \bibinfo {author} {\bibfnamefont {M.}~\bibnamefont {Bleicher}},\ and\
  \bibinfo {author} {\bibfnamefont {H.}~\bibnamefont {Stocker}},\ }\bibfield
  {title} {\bibinfo {title} {{A Fully Integrated Transport Approach to Heavy
  Ion Reactions with an Intermediate Hydrodynamic Stage}},\ }\href
  {https://doi.org/10.1103/PhysRevC.78.044901} {\bibfield  {journal} {\bibinfo
  {journal} {Phys. Rev. C}\ }\textbf {\bibinfo {volume} {78}},\ \bibinfo
  {pages} {044901} (\bibinfo {year} {2008})},\ \Eprint
  {https://arxiv.org/abs/0806.1695} {arXiv:0806.1695 [nucl-th]} \BibitemShut
  {NoStop}%
\bibitem [{\citenamefont {Shen}\ \emph {et~al.}(2016)\citenamefont {Shen},
  \citenamefont {Qiu}, \citenamefont {Song}, \citenamefont {Bernhard},
  \citenamefont {Bass},\ and\ \citenamefont {Heinz}}]{Shen:2014vra}%
  \BibitemOpen
  \bibfield  {author} {\bibinfo {author} {\bibfnamefont {C.}~\bibnamefont
  {Shen}}, \bibinfo {author} {\bibfnamefont {Z.}~\bibnamefont {Qiu}}, \bibinfo
  {author} {\bibfnamefont {H.}~\bibnamefont {Song}}, \bibinfo {author}
  {\bibfnamefont {J.}~\bibnamefont {Bernhard}}, \bibinfo {author}
  {\bibfnamefont {S.}~\bibnamefont {Bass}},\ and\ \bibinfo {author}
  {\bibfnamefont {U.}~\bibnamefont {Heinz}},\ }\bibfield  {title} {\bibinfo
  {title} {{The iEBE-VISHNU code package for relativistic heavy-ion
  collisions}},\ }\href {https://doi.org/10.1016/j.cpc.2015.08.039} {\bibfield
  {journal} {\bibinfo  {journal} {Comput. Phys. Commun.}\ }\textbf {\bibinfo
  {volume} {199}},\ \bibinfo {pages} {61} (\bibinfo {year} {2016})},\ \Eprint
  {https://arxiv.org/abs/1409.8164} {arXiv:1409.8164 [nucl-th]} \BibitemShut
  {NoStop}%
\bibitem [{\citenamefont {Bazavov}\ \emph {et~al.}(2014)\citenamefont {Bazavov}
  \emph {et~al.}}]{HotQCD:2014kol}%
  \BibitemOpen
  \bibfield  {author} {\bibinfo {author} {\bibfnamefont {A.}~\bibnamefont
  {Bazavov}} \emph {et~al.} (\bibinfo {collaboration} {HotQCD}),\ }\bibfield
  {title} {\bibinfo {title} {{Equation of state in ( 2+1 )-flavor QCD}},\
  }\href {https://doi.org/10.1103/PhysRevD.90.094503} {\bibfield  {journal}
  {\bibinfo  {journal} {Phys. Rev. D}\ }\textbf {\bibinfo {volume} {90}},\
  \bibinfo {pages} {094503} (\bibinfo {year} {2014})},\ \Eprint
  {https://arxiv.org/abs/1407.6387} {arXiv:1407.6387 [hep-lat]} \BibitemShut
  {NoStop}%
\bibitem [{\citenamefont {Cooper}\ and\ \citenamefont
  {Frye}(1974{\natexlab{a}})}]{PhysRevD.10.186}%
  \BibitemOpen
  \bibfield  {author} {\bibinfo {author} {\bibfnamefont {F.}~\bibnamefont
  {Cooper}}\ and\ \bibinfo {author} {\bibfnamefont {G.}~\bibnamefont {Frye}},\
  }\bibfield  {title} {\bibinfo {title} {Single-particle distribution in the
  hydrodynamic and statistical thermodynamic models of multiparticle
  production},\ }\href {https://doi.org/10.1103/PhysRevD.10.186} {\bibfield
  {journal} {\bibinfo  {journal} {Phys. Rev. D}\ }\textbf {\bibinfo {volume}
  {10}},\ \bibinfo {pages} {186} (\bibinfo {year}
  {1974}{\natexlab{a}})}\BibitemShut {NoStop}%
\bibitem [{\citenamefont {Rocha}\ \emph {et~al.}(2021)\citenamefont {Rocha},
  \citenamefont {Denicol},\ and\ \citenamefont {Noronha}}]{Rocha:2021zcw}%
  \BibitemOpen
  \bibfield  {author} {\bibinfo {author} {\bibfnamefont {G.~S.}\ \bibnamefont
  {Rocha}}, \bibinfo {author} {\bibfnamefont {G.~S.}\ \bibnamefont {Denicol}},\
  and\ \bibinfo {author} {\bibfnamefont {J.}~\bibnamefont {Noronha}},\
  }\bibfield  {title} {\bibinfo {title} {{Novel Relaxation Time Approximation
  to the Relativistic Boltzmann Equation}},\ }\href
  {https://doi.org/10.1103/PhysRevLett.127.042301} {\bibfield  {journal}
  {\bibinfo  {journal} {Phys. Rev. Lett.}\ }\textbf {\bibinfo {volume} {127}},\
  \bibinfo {pages} {042301} (\bibinfo {year} {2021})},\ \Eprint
  {https://arxiv.org/abs/2103.07489} {arXiv:2103.07489 [nucl-th]} \BibitemShut
  {NoStop}%
\bibitem [{\citenamefont {Rocha}\ and\ \citenamefont
  {Denicol}(2025)}]{Rocha:2025rkl}%
  \BibitemOpen
  \bibfield  {author} {\bibinfo {author} {\bibfnamefont {G.~S.}\ \bibnamefont
  {Rocha}}\ and\ \bibinfo {author} {\bibfnamefont {G.~S.}\ \bibnamefont
  {Denicol}},\ }\bibfield  {title} {\bibinfo {title} {{Relaxation time
  approximation for a multispecies relativistic gas}},\ }\href
  {https://doi.org/10.1103/gzc4-lnzd} {\bibfield  {journal} {\bibinfo
  {journal} {Phys. Rev. D}\ }\textbf {\bibinfo {volume} {112}},\ \bibinfo
  {pages} {076024} (\bibinfo {year} {2025})},\ \Eprint
  {https://arxiv.org/abs/2505.14823} {arXiv:2505.14823 [nucl-th]} \BibitemShut
  {NoStop}%
\bibitem [{\citenamefont {Borsanyi}\ \emph {et~al.}(2010)\citenamefont
  {Borsanyi}, \citenamefont {Endrodi}, \citenamefont {Fodor}, \citenamefont
  {Jakovac}, \citenamefont {Katz}, \citenamefont {Krieg}, \citenamefont
  {Ratti},\ and\ \citenamefont {Szabo}}]{Borsanyi:2010cj}%
  \BibitemOpen
  \bibfield  {author} {\bibinfo {author} {\bibfnamefont {S.}~\bibnamefont
  {Borsanyi}}, \bibinfo {author} {\bibfnamefont {G.}~\bibnamefont {Endrodi}},
  \bibinfo {author} {\bibfnamefont {Z.}~\bibnamefont {Fodor}}, \bibinfo
  {author} {\bibfnamefont {A.}~\bibnamefont {Jakovac}}, \bibinfo {author}
  {\bibfnamefont {S.~D.}\ \bibnamefont {Katz}}, \bibinfo {author}
  {\bibfnamefont {S.}~\bibnamefont {Krieg}}, \bibinfo {author} {\bibfnamefont
  {C.}~\bibnamefont {Ratti}},\ and\ \bibinfo {author} {\bibfnamefont {K.~K.}\
  \bibnamefont {Szabo}},\ }\bibfield  {title} {\bibinfo {title} {{The QCD
  equation of state with dynamical quarks}},\ }\href
  {https://doi.org/10.1007/JHEP11(2010)077} {\bibfield  {journal} {\bibinfo
  {journal} {JHEP}\ }\textbf {\bibinfo {volume} {11}},\ \bibinfo {pages}
  {077}},\ \Eprint {https://arxiv.org/abs/1007.2580} {arXiv:1007.2580
  [hep-lat]} \BibitemShut {NoStop}%
\bibitem [{\citenamefont {Bazavov}\ \emph {et~al.}(2009)\citenamefont {Bazavov}
  \emph {et~al.}}]{Bazavov:2009zn}%
  \BibitemOpen
  \bibfield  {author} {\bibinfo {author} {\bibfnamefont {A.}~\bibnamefont
  {Bazavov}} \emph {et~al.},\ }\bibfield  {title} {\bibinfo {title} {{Equation
  of state and QCD transition at finite temperature}},\ }\href
  {https://doi.org/10.1103/PhysRevD.80.014504} {\bibfield  {journal} {\bibinfo
  {journal} {Phys. Rev. D}\ }\textbf {\bibinfo {volume} {80}},\ \bibinfo
  {pages} {014504} (\bibinfo {year} {2009})},\ \Eprint
  {https://arxiv.org/abs/0903.4379} {arXiv:0903.4379 [hep-lat]} \BibitemShut
  {NoStop}%
\bibitem [{\citenamefont {Hagedorn}(1965)}]{Hagedorn:1965st}%
  \BibitemOpen
  \bibfield  {author} {\bibinfo {author} {\bibfnamefont {R.}~\bibnamefont
  {Hagedorn}},\ }\bibfield  {title} {\bibinfo {title} {{Statistical
  thermodynamics of strong interactions at high-energies}},\ }\href@noop {}
  {\bibfield  {journal} {\bibinfo  {journal} {Nuovo Cim. Suppl.}\ }\textbf
  {\bibinfo {volume} {3}},\ \bibinfo {pages} {147} (\bibinfo {year}
  {1965})}\BibitemShut {NoStop}%
\bibitem [{\citenamefont {Venugopalan}\ and\ \citenamefont
  {Prakash}(1992)}]{Venugopalan:1992hy}%
  \BibitemOpen
  \bibfield  {author} {\bibinfo {author} {\bibfnamefont {R.}~\bibnamefont
  {Venugopalan}}\ and\ \bibinfo {author} {\bibfnamefont {M.}~\bibnamefont
  {Prakash}},\ }\bibfield  {title} {\bibinfo {title} {{Thermal properties of
  interacting hadrons}},\ }\href {https://doi.org/10.1016/0375-9474(92)90005-5}
  {\bibfield  {journal} {\bibinfo  {journal} {Nucl. Phys. A}\ }\textbf
  {\bibinfo {volume} {546}},\ \bibinfo {pages} {718} (\bibinfo {year}
  {1992})}\BibitemShut {NoStop}%
\bibitem [{\citenamefont {Karsch}\ \emph {et~al.}(2003)\citenamefont {Karsch},
  \citenamefont {Redlich},\ and\ \citenamefont {Tawfik}}]{Karsch:2003vd}%
  \BibitemOpen
  \bibfield  {author} {\bibinfo {author} {\bibfnamefont {F.}~\bibnamefont
  {Karsch}}, \bibinfo {author} {\bibfnamefont {K.}~\bibnamefont {Redlich}},\
  and\ \bibinfo {author} {\bibfnamefont {A.}~\bibnamefont {Tawfik}},\
  }\bibfield  {title} {\bibinfo {title} {{Hadron resonance mass spectrum and
  lattice QCD thermodynamics}},\ }\href
  {https://doi.org/10.1140/epjc/s2003-01228-y} {\bibfield  {journal} {\bibinfo
  {journal} {Eur. Phys. J. C}\ }\textbf {\bibinfo {volume} {29}},\ \bibinfo
  {pages} {549} (\bibinfo {year} {2003})},\ \Eprint
  {https://arxiv.org/abs/hep-ph/0303108} {arXiv:hep-ph/0303108} \BibitemShut
  {NoStop}%
\bibitem [{\citenamefont {Huovinen}\ and\ \citenamefont
  {Petreczky}(2010)}]{Huovinen:2009yb}%
  \BibitemOpen
  \bibfield  {author} {\bibinfo {author} {\bibfnamefont {P.}~\bibnamefont
  {Huovinen}}\ and\ \bibinfo {author} {\bibfnamefont {P.}~\bibnamefont
  {Petreczky}},\ }\bibfield  {title} {\bibinfo {title} {{QCD Equation of State
  and Hadron Resonance Gas}},\ }\href
  {https://doi.org/10.1016/j.nuclphysa.2010.02.015} {\bibfield  {journal}
  {\bibinfo  {journal} {Nucl. Phys. A}\ }\textbf {\bibinfo {volume} {837}},\
  \bibinfo {pages} {26} (\bibinfo {year} {2010})},\ \Eprint
  {https://arxiv.org/abs/0912.2541} {arXiv:0912.2541 [hep-ph]} \BibitemShut
  {NoStop}%
\bibitem [{\citenamefont {Ratti}\ and\ \citenamefont
  {Bellwied}(2021)}]{Ratti:2021ubw}%
  \BibitemOpen
  \bibfield  {author} {\bibinfo {author} {\bibfnamefont {C.}~\bibnamefont
  {Ratti}}\ and\ \bibinfo {author} {\bibfnamefont {R.}~\bibnamefont
  {Bellwied}},\ }\href {https://doi.org/10.1007/978-3-030-67235-5} {\emph
  {\bibinfo {title} {{The Deconfinement Transition of QCD: Theory Meets
  Experiment}}}},\ \bibinfo {series} {Lecture Notes in Physics}, Vol.\ \bibinfo
  {volume} {981}\ (\bibinfo {year} {2021})\BibitemShut {NoStop}%
\bibitem [{\citenamefont {Aoki}\ \emph {et~al.}(2006)\citenamefont {Aoki},
  \citenamefont {Endrodi}, \citenamefont {Fodor}, \citenamefont {Katz},\ and\
  \citenamefont {Szabo}}]{Aoki:2006we}%
  \BibitemOpen
  \bibfield  {author} {\bibinfo {author} {\bibfnamefont {Y.}~\bibnamefont
  {Aoki}}, \bibinfo {author} {\bibfnamefont {G.}~\bibnamefont {Endrodi}},
  \bibinfo {author} {\bibfnamefont {Z.}~\bibnamefont {Fodor}}, \bibinfo
  {author} {\bibfnamefont {S.~D.}\ \bibnamefont {Katz}},\ and\ \bibinfo
  {author} {\bibfnamefont {K.~K.}\ \bibnamefont {Szabo}},\ }\bibfield  {title}
  {\bibinfo {title} {{The Order of the quantum chromodynamics transition
  predicted by the standard model of particle physics}},\ }\href
  {https://doi.org/10.1038/nature05120} {\bibfield  {journal} {\bibinfo
  {journal} {Nature}\ }\textbf {\bibinfo {volume} {443}},\ \bibinfo {pages}
  {675} (\bibinfo {year} {2006})},\ \Eprint
  {https://arxiv.org/abs/hep-lat/0611014} {arXiv:hep-lat/0611014} \BibitemShut
  {NoStop}%
\bibitem [{\citenamefont {Cooper}\ and\ \citenamefont
  {Frye}(1974{\natexlab{b}})}]{Cooper:1974mv}%
  \BibitemOpen
  \bibfield  {author} {\bibinfo {author} {\bibfnamefont {F.}~\bibnamefont
  {Cooper}}\ and\ \bibinfo {author} {\bibfnamefont {G.}~\bibnamefont {Frye}},\
  }\bibfield  {title} {\bibinfo {title} {{Comment on the Single Particle
  Distribution in the Hydrodynamic and Statistical Thermodynamic Models of
  Multiparticle Production}},\ }\href {https://doi.org/10.1103/PhysRevD.10.186}
  {\bibfield  {journal} {\bibinfo  {journal} {Phys. Rev. D}\ }\textbf {\bibinfo
  {volume} {10}},\ \bibinfo {pages} {186} (\bibinfo {year}
  {1974}{\natexlab{b}})}\BibitemShut {NoStop}%
\bibitem [{\citenamefont {de~Groot}\ \emph {et~al.}(1980)\citenamefont
  {de~Groot}, \citenamefont {van Leeuwen},\ and\ \citenamefont {van
  Weert}}]{deGroot:80relativistic}%
  \BibitemOpen
  \bibfield  {author} {\bibinfo {author} {\bibfnamefont {S.}~\bibnamefont
  {de~Groot}}, \bibinfo {author} {\bibfnamefont {W.}~\bibnamefont {van
  Leeuwen}},\ and\ \bibinfo {author} {\bibfnamefont {C.}~\bibnamefont {van
  Weert}},\ }\href@noop {} {\emph {\bibinfo {title} {Relativistic Kinetic
  Theory: Principles and Applications}}}\ (\bibinfo  {publisher} {North-Holland
  Publishing Co.},\ \bibinfo {year} {1980})\BibitemShut {NoStop}%
\bibitem [{\citenamefont {Landau}\ and\ \citenamefont
  {Lifshitz}(1987)}]{LandauLifshitzFluids}%
  \BibitemOpen
  \bibfield  {author} {\bibinfo {author} {\bibfnamefont {L.~D.}\ \bibnamefont
  {Landau}}\ and\ \bibinfo {author} {\bibfnamefont {E.~M.}\ \bibnamefont
  {Lifshitz}},\ }\href@noop {} {\emph {\bibinfo {title} {Fluid Mechanics -
  Volume 6 (Corse of Theoretical Physics)}}},\ \bibinfo {edition} {2nd}\ ed.\
  (\bibinfo  {publisher} {Butterworth-Heinemann},\ \bibinfo {year} {1987})\ p.\
  \bibinfo {pages} {552}\BibitemShut {NoStop}%
\bibitem [{\citenamefont {Chapman}(1916)}]{chapman1916vi}%
  \BibitemOpen
  \bibfield  {author} {\bibinfo {author} {\bibfnamefont {S.}~\bibnamefont
  {Chapman}},\ }\bibfield  {title} {\bibinfo {title} {Vi. on the law of
  distribution of molecular velocities, and on the theory of viscosity and
  thermal conduction, in a non-uniform simple monatomic gas},\ }\href@noop {}
  {\bibfield  {journal} {\bibinfo  {journal} {Philosophical Transactions of the
  Royal Society of London. Series A, Containing Papers of a Mathematical or
  Physical Character}\ }\textbf {\bibinfo {volume} {216}},\ \bibinfo {pages}
  {279} (\bibinfo {year} {1916})}\BibitemShut {NoStop}%
\bibitem [{\citenamefont {Enskog}(1917)}]{enskog1917kinetische}%
  \BibitemOpen
  \bibfield  {author} {\bibinfo {author} {\bibfnamefont {D.}~\bibnamefont
  {Enskog}},\ }\href@noop {} {\bibinfo {title} {Kinetische theorie der
  {V}org{\"a}nge in m{\"a}ssig verd{\"u}nnten gasen. {I}. {A}llgemeiner teil}}
  (\bibinfo {year} {1917})\BibitemShut {NoStop}%
\bibitem [{\citenamefont {Cercignani}\ and\ \citenamefont
  {Kremer}(2002)}]{cercignani:02relativistic}%
  \BibitemOpen
  \bibfield  {author} {\bibinfo {author} {\bibfnamefont {C.}~\bibnamefont
  {Cercignani}}\ and\ \bibinfo {author} {\bibfnamefont {G.~M.}\ \bibnamefont
  {Kremer}},\ }\href@noop {} {\emph {\bibinfo {title} {The Relativistic
  {B}oltzmann Equation: Theory and Applications}}}\ (\bibinfo  {publisher}
  {Springer},\ \bibinfo {year} {2002})\BibitemShut {NoStop}%
\bibitem [{\citenamefont {Denicol}\ and\ \citenamefont
  {Rischke}(2021)}]{Denicol:2021}%
  \BibitemOpen
  \bibfield  {author} {\bibinfo {author} {\bibfnamefont {G.}~\bibnamefont
  {Denicol}}\ and\ \bibinfo {author} {\bibfnamefont {D.~H.}\ \bibnamefont
  {Rischke}},\ }\href@noop {} {\emph {\bibinfo {title} {Microscopic Foundations
  of Relativistic Fluid Dynamics}}}\ (\bibinfo  {publisher} {Springer},\
  \bibinfo {year} {2021})\BibitemShut {NoStop}%
\bibitem [{\citenamefont {Rocha}\ \emph {et~al.}(2022)\citenamefont {Rocha},
  \citenamefont {Denicol},\ and\ \citenamefont {Noronha}}]{Rocha:2022ind}%
  \BibitemOpen
  \bibfield  {author} {\bibinfo {author} {\bibfnamefont {G.~S.}\ \bibnamefont
  {Rocha}}, \bibinfo {author} {\bibfnamefont {G.~S.}\ \bibnamefont {Denicol}},\
  and\ \bibinfo {author} {\bibfnamefont {J.}~\bibnamefont {Noronha}},\
  }\bibfield  {title} {\bibinfo {title} {{Perturbative approaches in
  relativistic kinetic theory and the emergence of first-order
  hydrodynamics}},\ }\href {https://doi.org/10.1103/PhysRevD.106.036010}
  {\bibfield  {journal} {\bibinfo  {journal} {Phys. Rev. D}\ }\textbf {\bibinfo
  {volume} {106}},\ \bibinfo {pages} {036010} (\bibinfo {year} {2022})},\
  \Eprint {https://arxiv.org/abs/2205.00078} {arXiv:2205.00078 [nucl-th]}
  \BibitemShut {NoStop}%
\bibitem [{\citenamefont {Rocha}\ \emph {et~al.}(2024)\citenamefont {Rocha},
  \citenamefont {Wagner}, \citenamefont {Denicol}, \citenamefont {Noronha},\
  and\ \citenamefont {Rischke}}]{Rocha:2023ilf}%
  \BibitemOpen
  \bibfield  {author} {\bibinfo {author} {\bibfnamefont {G.~S.}\ \bibnamefont
  {Rocha}}, \bibinfo {author} {\bibfnamefont {D.}~\bibnamefont {Wagner}},
  \bibinfo {author} {\bibfnamefont {G.~S.}\ \bibnamefont {Denicol}}, \bibinfo
  {author} {\bibfnamefont {J.}~\bibnamefont {Noronha}},\ and\ \bibinfo {author}
  {\bibfnamefont {D.~H.}\ \bibnamefont {Rischke}},\ }\bibfield  {title}
  {\bibinfo {title} {{Theories of Relativistic Dissipative Fluid Dynamics}},\
  }\href {https://doi.org/10.3390/e26030189} {\bibfield  {journal} {\bibinfo
  {journal} {Entropy}\ }\textbf {\bibinfo {volume} {26}},\ \bibinfo {pages}
  {189} (\bibinfo {year} {2024})},\ \Eprint {https://arxiv.org/abs/2311.15063}
  {arXiv:2311.15063 [nucl-th]} \BibitemShut {NoStop}%
\bibitem [{\citenamefont {McNelis}\ \emph {et~al.}(2021)\citenamefont
  {McNelis}, \citenamefont {Everett},\ and\ \citenamefont
  {Heinz}}]{McNelis:2019auj}%
  \BibitemOpen
  \bibfield  {author} {\bibinfo {author} {\bibfnamefont {M.}~\bibnamefont
  {McNelis}}, \bibinfo {author} {\bibfnamefont {D.}~\bibnamefont {Everett}},\
  and\ \bibinfo {author} {\bibfnamefont {U.}~\bibnamefont {Heinz}},\ }\bibfield
   {title} {\bibinfo {title} {{Particlization in fluid dynamical simulations of
  heavy-ion collisions: The i S3D module}},\ }\href
  {https://doi.org/10.1016/j.cpc.2020.107604} {\bibfield  {journal} {\bibinfo
  {journal} {Comput. Phys. Commun.}\ }\textbf {\bibinfo {volume} {258}},\
  \bibinfo {pages} {107604} (\bibinfo {year} {2021})},\ \Eprint
  {https://arxiv.org/abs/1912.08271} {arXiv:1912.08271 [nucl-th]} \BibitemShut
  {NoStop}%
\bibitem [{\citenamefont {Denicol}\ and\ \citenamefont
  {Noronha}(2024)}]{Denicol:2022bsq}%
  \BibitemOpen
  \bibfield  {author} {\bibinfo {author} {\bibfnamefont {G.~S.}\ \bibnamefont
  {Denicol}}\ and\ \bibinfo {author} {\bibfnamefont {J.}~\bibnamefont
  {Noronha}},\ }\bibfield  {title} {\bibinfo {title} {{Spectrum of the
  Boltzmann collision operator for {\ensuremath{\lambda}}{\ensuremath{\phi}}4
  theory in the classical regime}},\ }\href
  {https://doi.org/10.1016/j.physletb.2024.138487} {\bibfield  {journal}
  {\bibinfo  {journal} {Phys. Lett. B}\ }\textbf {\bibinfo {volume} {850}},\
  \bibinfo {pages} {138487} (\bibinfo {year} {2024})},\ \Eprint
  {https://arxiv.org/abs/2209.10370} {arXiv:2209.10370 [nucl-th]} \BibitemShut
  {NoStop}%
\bibitem [{\citenamefont {Anderson}\ and\ \citenamefont
  {Witting}(1974)}]{ANDERSON1974489}%
  \BibitemOpen
  \bibfield  {author} {\bibinfo {author} {\bibfnamefont {J.}~\bibnamefont
  {Anderson}}\ and\ \bibinfo {author} {\bibfnamefont {H.}~\bibnamefont
  {Witting}},\ }\bibfield  {title} {\bibinfo {title} {Relativistic quantum
  transport coefficients},\ }\href
  {https://doi.org/https://doi.org/10.1016/0031-8914(74)90356-5} {\bibfield
  {journal} {\bibinfo  {journal} {Physica}\ }\textbf {\bibinfo {volume} {74}},\
  \bibinfo {pages} {489} (\bibinfo {year} {1974})}\BibitemShut {NoStop}%
\bibitem [{\citenamefont {Rocha}\ \emph {et~al.}(2023)\citenamefont {Rocha},
  \citenamefont {Denicol}, \citenamefont {Ferreira},\ and\ \citenamefont
  {Noronha}}]{Rocha:2022crt}%
  \BibitemOpen
  \bibfield  {author} {\bibinfo {author} {\bibfnamefont {G.~S.}\ \bibnamefont
  {Rocha}}, \bibinfo {author} {\bibfnamefont {G.~S.}\ \bibnamefont {Denicol}},
  \bibinfo {author} {\bibfnamefont {M.~N.}\ \bibnamefont {Ferreira}},\ and\
  \bibinfo {author} {\bibfnamefont {J.}~\bibnamefont {Noronha}},\ }\bibfield
  {title} {\bibinfo {title} {{Novel Relaxation Time Approximation: A Consistent
  Calculation of Transport Coefficients with QCD-inspired Relaxation Times}},\
  }\href {https://doi.org/10.5506/APhysPolBSupp.16.1-A29} {\bibfield  {journal}
  {\bibinfo  {journal} {Acta Phys. Polon. Supp.}\ }\textbf {\bibinfo {volume}
  {16}},\ \bibinfo {pages} {1} (\bibinfo {year} {2023})},\ \Eprint
  {https://arxiv.org/abs/2207.11286} {arXiv:2207.11286 [nucl-th]} \BibitemShut
  {NoStop}%
\bibitem [{\citenamefont {Dash}\ \emph {et~al.}(2022)\citenamefont {Dash},
  \citenamefont {Bhadury}, \citenamefont {Jaiswal},\ and\ \citenamefont
  {Jaiswal}}]{Dash:2021ibx}%
  \BibitemOpen
  \bibfield  {author} {\bibinfo {author} {\bibfnamefont {D.}~\bibnamefont
  {Dash}}, \bibinfo {author} {\bibfnamefont {S.}~\bibnamefont {Bhadury}},
  \bibinfo {author} {\bibfnamefont {S.}~\bibnamefont {Jaiswal}},\ and\ \bibinfo
  {author} {\bibfnamefont {A.}~\bibnamefont {Jaiswal}},\ }\bibfield  {title}
  {\bibinfo {title} {{Extended relaxation time approximation and relativistic
  dissipative hydrodynamics}},\ }\href
  {https://doi.org/10.1016/j.physletb.2022.137202} {\bibfield  {journal}
  {\bibinfo  {journal} {Phys. Lett. B}\ }\textbf {\bibinfo {volume} {831}},\
  \bibinfo {pages} {137202} (\bibinfo {year} {2022})},\ \Eprint
  {https://arxiv.org/abs/2112.14581} {arXiv:2112.14581 [nucl-th]} \BibitemShut
  {NoStop}%
\bibitem [{\citenamefont {Calzetta}\ and\ \citenamefont
  {Hu}(1988)}]{Calzetta:1986cq}%
  \BibitemOpen
  \bibfield  {author} {\bibinfo {author} {\bibfnamefont {E.}~\bibnamefont
  {Calzetta}}\ and\ \bibinfo {author} {\bibfnamefont {B.~L.}\ \bibnamefont
  {Hu}},\ }\bibfield  {title} {\bibinfo {title} {{Nonequilibrium Quantum
  Fields: Closed Time Path Effective Action, Wigner Function and Boltzmann
  Equation}},\ }\href {https://doi.org/10.1103/PhysRevD.37.2878} {\bibfield
  {journal} {\bibinfo  {journal} {Phys. Rev. D}\ }\textbf {\bibinfo {volume}
  {37}},\ \bibinfo {pages} {2878} (\bibinfo {year} {1988})}\BibitemShut
  {NoStop}%
\bibitem [{\citenamefont {Bass}\ and\ \citenamefont {{et
  al}}(1998)}]{Bass:1998ca}%
  \BibitemOpen
  \bibfield  {author} {\bibinfo {author} {\bibfnamefont {S.~A.}\ \bibnamefont
  {Bass}}\ and\ \bibinfo {author} {\bibnamefont {{et al}}},\ }\href@noop {}
  {\bibfield  {journal} {\bibinfo  {journal} {Prog. Part. Nucl. Phys.}\
  }\textbf {\bibinfo {volume} {41}},\ \bibinfo {pages} {255} (\bibinfo {year}
  {1998})}\BibitemShut {NoStop}%
\bibitem [{\citenamefont {Bleicher}\ and\ \citenamefont {{et
  al}}(1999)}]{Bleicher:1999xi}%
  \BibitemOpen
  \bibfield  {author} {\bibinfo {author} {\bibfnamefont {M.}~\bibnamefont
  {Bleicher}}\ and\ \bibinfo {author} {\bibnamefont {{et al}}},\ }\href@noop {}
  {\bibfield  {journal} {\bibinfo  {journal} {J. Phys.}\ }\textbf {\bibinfo
  {volume} {G25}},\ \bibinfo {pages} {1859} (\bibinfo {year}
  {1999})}\BibitemShut {NoStop}%
\bibitem [{\citenamefont {Moreland}\ \emph {et~al.}(2015)\citenamefont
  {Moreland}, \citenamefont {Bernhard},\ and\ \citenamefont
  {Bass}}]{Moreland:2014oya}%
  \BibitemOpen
  \bibfield  {author} {\bibinfo {author} {\bibfnamefont {J.~S.}\ \bibnamefont
  {Moreland}}, \bibinfo {author} {\bibfnamefont {J.~E.}\ \bibnamefont
  {Bernhard}},\ and\ \bibinfo {author} {\bibfnamefont {S.~A.}\ \bibnamefont
  {Bass}},\ }\href@noop {} {\bibfield  {journal} {\bibinfo  {journal} {Phys.
  Rev.}\ }\textbf {\bibinfo {volume} {C92}},\ \bibinfo {pages} {011901}
  (\bibinfo {year} {2015})}\BibitemShut {NoStop}%
\bibitem [{\citenamefont {Broniowski}\ \emph {et~al.}(2009)\citenamefont
  {Broniowski}, \citenamefont {Florkowski}, \citenamefont {Chojnacki},\ and\
  \citenamefont {Kisiel}}]{Broniowski:2008qk}%
  \BibitemOpen
  \bibfield  {author} {\bibinfo {author} {\bibfnamefont {W.}~\bibnamefont
  {Broniowski}}, \bibinfo {author} {\bibfnamefont {W.}~\bibnamefont
  {Florkowski}}, \bibinfo {author} {\bibfnamefont {M.}~\bibnamefont
  {Chojnacki}},\ and\ \bibinfo {author} {\bibfnamefont {A.}~\bibnamefont
  {Kisiel}},\ }\bibfield  {title} {\bibinfo {title} {{Free-streaming
  approximation in early dynamics of relativistic heavy-ion collisions}},\
  }\href {https://doi.org/10.1103/PhysRevC.80.034902} {\bibfield  {journal}
  {\bibinfo  {journal} {Phys. Rev.}\ }\textbf {\bibinfo {volume} {C80}},\
  \bibinfo {pages} {034902} (\bibinfo {year} {2009})}\BibitemShut {NoStop}%
\bibitem [{\citenamefont {Liu}\ \emph {et~al.}(2015)\citenamefont {Liu},
  \citenamefont {Shen},\ and\ \citenamefont {Heinz}}]{Liu:2015nwa}%
  \BibitemOpen
  \bibfield  {author} {\bibinfo {author} {\bibfnamefont {J.}~\bibnamefont
  {Liu}}, \bibinfo {author} {\bibfnamefont {C.}~\bibnamefont {Shen}},\ and\
  \bibinfo {author} {\bibfnamefont {U.}~\bibnamefont {Heinz}},\ }\bibfield
  {title} {\bibinfo {title} {{Pre-equilibrium evolution effects on heavy-ion
  collision observables}},\ }\href {https://doi.org/10.1103/PhysRevC.92.049904,
  10.1103/PhysRevC.91.064906} {\bibfield  {journal} {\bibinfo  {journal} {Phys.
  Rev.}\ }\textbf {\bibinfo {volume} {C91}},\ \bibinfo {pages} {064906}
  (\bibinfo {year} {2015})},\ \bibinfo {note} {[Erratum: Phys.
  Rev.C92,049904(2015)]}\BibitemShut {NoStop}%
\bibitem [{\citenamefont {Schenke}\ \emph {et~al.}(2010)\citenamefont
  {Schenke}, \citenamefont {Jeon},\ and\ \citenamefont
  {Gale}}]{Schenke:2010nt}%
  \BibitemOpen
  \bibfield  {author} {\bibinfo {author} {\bibfnamefont {B.}~\bibnamefont
  {Schenke}}, \bibinfo {author} {\bibfnamefont {S.}~\bibnamefont {Jeon}},\ and\
  \bibinfo {author} {\bibfnamefont {C.}~\bibnamefont {Gale}},\ }\href@noop {}
  {\bibfield  {journal} {\bibinfo  {journal} {Phys. Rev.}\ }\textbf {\bibinfo
  {volume} {C82}},\ \bibinfo {pages} {014903} (\bibinfo {year}
  {2010})}\BibitemShut {NoStop}%
\bibitem [{\citenamefont {Schenke}\ \emph {et~al.}(2012)\citenamefont
  {Schenke}, \citenamefont {Jeon},\ and\ \citenamefont
  {Gale}}]{Schenke:2011bn}%
  \BibitemOpen
  \bibfield  {author} {\bibinfo {author} {\bibfnamefont {B.}~\bibnamefont
  {Schenke}}, \bibinfo {author} {\bibfnamefont {S.}~\bibnamefont {Jeon}},\ and\
  \bibinfo {author} {\bibfnamefont {C.}~\bibnamefont {Gale}},\ }\href@noop {}
  {\bibfield  {journal} {\bibinfo  {journal} {Phys. Rev.}\ }\textbf {\bibinfo
  {volume} {C85}},\ \bibinfo {pages} {024901} (\bibinfo {year}
  {2012})}\BibitemShut {NoStop}%
\bibitem [{\citenamefont {Ryu}\ \emph {et~al.}(2015)\citenamefont {Ryu},
  \citenamefont {Paquet}, \citenamefont {Shen}, \citenamefont {Denicol},
  \citenamefont {Schenke}, \citenamefont {Jeon},\ and\ \citenamefont
  {Gale}}]{Ryu:2015vwa}%
  \BibitemOpen
  \bibfield  {author} {\bibinfo {author} {\bibfnamefont {S.}~\bibnamefont
  {Ryu}}, \bibinfo {author} {\bibfnamefont {J.~F.}\ \bibnamefont {Paquet}},
  \bibinfo {author} {\bibfnamefont {C.}~\bibnamefont {Shen}}, \bibinfo {author}
  {\bibfnamefont {G.}~\bibnamefont {Denicol}}, \bibinfo {author} {\bibfnamefont
  {B.}~\bibnamefont {Schenke}}, \bibinfo {author} {\bibfnamefont
  {S.}~\bibnamefont {Jeon}},\ and\ \bibinfo {author} {\bibfnamefont
  {C.}~\bibnamefont {Gale}},\ }\bibfield  {title} {\bibinfo {title}
  {{Importance of the Bulk Viscosity of QCD in Ultrarelativistic Heavy-Ion
  Collisions}},\ }\href {https://doi.org/10.1103/PhysRevLett.115.132301}
  {\bibfield  {journal} {\bibinfo  {journal} {Phys. Rev. Lett.}\ }\textbf
  {\bibinfo {volume} {115}},\ \bibinfo {pages} {132301} (\bibinfo {year}
  {2015})},\ \Eprint {https://arxiv.org/abs/1502.01675} {arXiv:1502.01675
  [nucl-th]} \BibitemShut {NoStop}%
\bibitem [{\citenamefont {Paquet}\ \emph {et~al.}(2016)\citenamefont {Paquet},
  \citenamefont {Shen}, \citenamefont {Denicol}, \citenamefont {Luzum},
  \citenamefont {Schenke}, \citenamefont {Jeon},\ and\ \citenamefont
  {Gale}}]{Paquet:2015lta}%
  \BibitemOpen
  \bibfield  {author} {\bibinfo {author} {\bibfnamefont {J.-F.}\ \bibnamefont
  {Paquet}}, \bibinfo {author} {\bibfnamefont {C.}~\bibnamefont {Shen}},
  \bibinfo {author} {\bibfnamefont {G.~S.}\ \bibnamefont {Denicol}}, \bibinfo
  {author} {\bibfnamefont {M.}~\bibnamefont {Luzum}}, \bibinfo {author}
  {\bibfnamefont {B.}~\bibnamefont {Schenke}}, \bibinfo {author} {\bibfnamefont
  {S.}~\bibnamefont {Jeon}},\ and\ \bibinfo {author} {\bibfnamefont
  {C.}~\bibnamefont {Gale}},\ }\bibfield  {title} {\bibinfo {title}
  {{Production of photons in relativistic heavy-ion collisions}},\ }\href
  {https://doi.org/10.1103/PhysRevC.93.044906} {\bibfield  {journal} {\bibinfo
  {journal} {Phys. Rev. C}\ }\textbf {\bibinfo {volume} {93}},\ \bibinfo
  {pages} {044906} (\bibinfo {year} {2016})},\ \Eprint
  {https://arxiv.org/abs/1509.06738} {arXiv:1509.06738 [hep-ph]} \BibitemShut
  {NoStop}%
\bibitem [{\citenamefont {Everett}\ \emph {et~al.}(2021)\citenamefont {Everett}
  \emph {et~al.}}]{JETSCAPE:2020mzn}%
  \BibitemOpen
  \bibfield  {author} {\bibinfo {author} {\bibfnamefont {D.}~\bibnamefont
  {Everett}} \emph {et~al.} (\bibinfo {collaboration} {JETSCAPE}),\ }\bibfield
  {title} {\bibinfo {title} {{Multisystem Bayesian constraints on the transport
  coefficients of QCD matter}},\ }\href
  {https://doi.org/10.1103/PhysRevC.103.054904} {\bibfield  {journal} {\bibinfo
   {journal} {Phys. Rev. C}\ }\textbf {\bibinfo {volume} {103}},\ \bibinfo
  {pages} {054904} (\bibinfo {year} {2021})},\ \Eprint
  {https://arxiv.org/abs/2011.01430} {arXiv:2011.01430 [hep-ph]} \BibitemShut
  {NoStop}%
\bibitem [{\citenamefont {Moreland}\ \emph {et~al.}(2020)\citenamefont
  {Moreland}, \citenamefont {Bernhard},\ and\ \citenamefont
  {Bass}}]{Moreland:2018gsh}%
  \BibitemOpen
  \bibfield  {author} {\bibinfo {author} {\bibfnamefont {J.~S.}\ \bibnamefont
  {Moreland}}, \bibinfo {author} {\bibfnamefont {J.~E.}\ \bibnamefont
  {Bernhard}},\ and\ \bibinfo {author} {\bibfnamefont {S.~A.}\ \bibnamefont
  {Bass}},\ }\bibfield  {title} {\bibinfo {title} {{Bayesian calibration of a
  hybrid nuclear collision model using p-Pb and Pb-Pb data at energies
  available at the CERN Large Hadron Collider}},\ }\href
  {https://doi.org/10.1103/PhysRevC.101.024911} {\bibfield  {journal} {\bibinfo
   {journal} {Phys. Rev. C}\ }\textbf {\bibinfo {volume} {101}},\ \bibinfo
  {pages} {024911} (\bibinfo {year} {2020})},\ \Eprint
  {https://arxiv.org/abs/1808.02106} {arXiv:1808.02106 [nucl-th]} \BibitemShut
  {NoStop}%
\bibitem [{\citenamefont {Weller}\ and\ \citenamefont
  {Romatschke}(2017)}]{Weller:2017tsr}%
  \BibitemOpen
  \bibfield  {author} {\bibinfo {author} {\bibfnamefont {R.~D.}\ \bibnamefont
  {Weller}}\ and\ \bibinfo {author} {\bibfnamefont {P.}~\bibnamefont
  {Romatschke}},\ }\bibfield  {title} {\bibinfo {title} {{One fluid to rule
  them all: viscous hydrodynamic description of event-by-event central p+p,
  p+Pb and Pb+Pb collisions at $\sqrt{s}=5.02$ TeV}},\ }\href
  {https://doi.org/10.1016/j.physletb.2017.09.077} {\bibfield  {journal}
  {\bibinfo  {journal} {Phys. Lett. B}\ }\textbf {\bibinfo {volume} {774}},\
  \bibinfo {pages} {351} (\bibinfo {year} {2017})},\ \Eprint
  {https://arxiv.org/abs/1701.07145} {arXiv:1701.07145 [nucl-th]} \BibitemShut
  {NoStop}%
\end{thebibliography}%

\end{document}